\documentclass[reqno,12pt]{article}
\usepackage{amsfonts,amssymb,amsmath,amscd,bm,cite,delarray,subfigure}
\usepackage[dvips]{color}
\usepackage[dvips]{graphicx}

\numberwithin{equation}{section}

\font\capital=rsfs11
\font\scriptcapital=rsfs10 at 7 truept
\font\scriptscriptcapital=rsfs10 at 5 truept
\def\scri{\fam=15}

\font\sansserif=cmss12
\font\scriptsansserif=cmss12 at 7 truept
\font\scriptscriptsansserif=cmss12 at 5 truept
\font\euler=eusm10
\font\scripteuler=eusm7
\font\scriptscripteuler=eusm5 
\newcommand{\barpartial}{\overline{\partial}}
\newcommand{\barnabla}{\overline{\nabla}}
\newcommand{\ul}[1]{{\underline{#1}}{}}

\begin{document}

\hrule\vskip.4cm
\hbox to 14.5 truecm{November 2005 \hfil DFUB 05--11}
\hbox to 14.5 truecm{Version 1  \hfil hep-th/0511144}
\vskip.4cm\hrule
\vskip.7cm
\begin{large}
\centerline{\bf A TOPOLOGICAL SIGMA MODEL}   
\centerline{\bf OF BIKAEHELR GEOMETRY}   
\end{large}
\vskip.2cm
\centerline{by}
\vskip.2cm
\centerline{\bf Roberto Zucchini}
\centerline{\it Dipartimento di Fisica, Universit\`a degli Studi di Bologna}
\centerline{\it V. Irnerio 46, I-40126 Bologna, Italy}
\centerline{\it I.N.F.N., sezione di Bologna, Italy}
\centerline{\it E--mail: zucchinir@bo.infn.it}
\vskip.7cm
\hrule
\vskip.7cm
\centerline{\bf Abstract} 
\par\noindent
BiKaehler geometry is characterized by a Riemannian metric $g_{ab}$ and two 
covariantly constant generally non commuting complex structures $K_\pm{}^a{}_b$,
with respect to which $g_{ab}$ is Hermitian. It is a particular case of the 
biHermitian geometry of Gates, Hull and Roceck, the most general sigma model 
target space geometry allowing for $(2,2)$ world sheet supersymmetry. 
We present a sigma model for biKaehler geometry that is topological in the
following sense: $i$) the action is invariant under a fermionic 
symmetry $\delta$; $ii$) $\delta$ is nilpotent 
on shell; $iii$) the action is $\delta$--exact on shell up to a topological term;
$iv$) the resulting field theory depends only on a subset of the target space
geometrical data. 
The biKaehler sigma model is obtainable by gauge fixing the Hitchin model with
generalized Kaehler target space. It further contains the customary $A$ 
topological sigma model as a particular case. However, it is not seemingly 
related to the $(2,2)$ supersymmetric biKaehler sigma model by twisting in general. 
\par\noindent
Keywords: Topological Field Theory, Generalized Complex Geometry.

\vfill\eject

\begin{small}
\section{\bf Introduction}
\label{sec:intro}
\end{small}

Type II superstring Calabi--Yau compactifications are described by $(2,2)$ 
superconformal sigma models with Calabi--Yau target manifolds. These field theories
are however rather complicated and, so, they are difficult to study. 
In 1988, Witten showed that a $(2,2)$ supersymmetric sigma model on a 
Calabi--Yau space could be twisted in two different ways,   
to give the so called $A$ and $B$ topological sigma models \cite{Witten1,Witten2}.  
Unlike the original untwisted sigma model, the topological models are soluble:
the calculation of observables can be reduced to classical problems of geometry.
For this reason, the topological sigma models constitute an ideal field theoretic ground
for the in depth study of 2--dimensional supersymmetric field theories. 

Witten's analysis was restricted to the case where the sigma model target space geometry 
was Kaehler. In a classic paper, Gates, Hull and Roceck 
\cite{Gates} showed that, for a 2--dimensional sigma model, the most general
target space geometry allowing for $(2,2)$ supersymmetry 
was biHermitian or Kaehler with torsion geometry. This is characterized by 
a Riemannian metric $g_{ab}$, two generally non commuting complex structures 
$K_\pm{}^a{}_b$ and a closed $3$--form $H_{abc}$,
such that $g_{ab}$ is Hermitian with respect to both the $K_\pm{}^a{}_b$ and 
the $K_\pm{}^a{}_b$ are parallel with respect to two different metric
connections with torsion proportional to $\pm H_{abc}$ 
\cite{Rocek,Ivanov,Bogaerts,Lyakhovich}.
This geometry is much more general than that considered by Witten, which
corresponds to the case where $K_+{}^a{}_b=\pm K_-{}^a{}_b$ and $H_{abc}=0$.

In 2002, Hitchin formulated the notion of generalized complex geometry, which
at the same time unifies and extends the customary notions of complex and 
symplectic geometry and incorporates a natural generalization of Calabi--Yau 
geometry \cite{Hitchin}. Hitchin's ideas were developed by Gualtieri \cite{Gualtieri},
who introduced the notion of generalized Kaehler geometry and showed that 
the biHermitian geometry of Gates, Hull and Roceck was equivalent to the
latter. 

In refs. \cite{Kapustin1, Kapustin2}, Kapustin and Kapustin and Li defined 
and studied the analogues of $A$ and $B$ models for $(2,2)$ supersymmetric
sigma models with $H$ field and showed that the results were
naturally expressed in the language of 
generalized complex geometry. Simultaneously, other attempts were made to 
construct sigma models based on generalized 
complex or Kaehler geometry, by invoking world sheet supersymmetry, 
employing the Batalin--Vilkovisky quantization algorithm, etc. 
\cite{Lindstrom1,Lindstrom2,Lindstrom3,Lindstrom4,Bredthauer1,
Zabzine1,Zabzine2,Zucchini1,Zucchini2,Chiantese}. 
All these attempts were somehow unsatisfactory either because they
remained confined to the analysis of geometrical aspects of the sigma
models or because they yielded field theories, which though interesting in
their own, were not directly suitable for quantization and showed no apparent
kinship with Witten's $A$ and $B$ models. 

In this paper, we present a topological sigma model with target space biKaehler
geometry. This geometry is characterized by a Riemannian metric $g_{ab}$ 
and two covariantly constant generally non commuting complex structures $K_\pm{}^a{}_b$,
with respect to which $g_{ab}$ is Hermitian. It is a particular case of the 
biHermitian geometry of ref. \cite{Gates} corresponding to $H_{abc}=0$.

The biKaehler sigma model expondeded in the paper has all the basic 
features of a topological sigma model as summarized below.

\par\noindent
{\it a}. The action $S$ possesses an odd symmetry $\delta$. 

\par\noindent
{\it b}. $\delta$ is nilpotent on shell.

\par\noindent
{\it c}. $S$ is $\delta$--exact on shell up to a topological term, 
with some weak restrictions on the target space geometry.

\par\noindent
{\it d}. The resulting field theory depends only on a certain combination of 
the target space geometrical data. 
 
\par\noindent
The model also  has other interesting features.

\par\noindent
{\it e}. It is obtainable by gauge fixing the Hitchin model
\cite{Zucchini1,Zucchini2} with generalized Kaehler target space 
\cite{Gualtieri} according to the general philosophy of 
Alexandrov, Kontsevich, Schwartz and Zaboronsky \cite{AKSZ}.

\par\noindent
{\it f}. In the particular case $K_+{}^a{}_b=-K_-{}^a{}_b$, 
it reproduces Witten's $A$ topological sigma model \cite{Witten1,Witten2}.   
It also yields topological sigma models 
for product structure and hyperKaehler target space geometries. 

Roughly speaking, the field content of the model consists of the fields of the 
$A$ topological sigma model plus a further 1-form field. This latter
becomes non propagating and decouples in the particular case of the 
$A$ model, but it does not in the general case. 
Also, the algebra of local topological observables is isomorphic to the
Poisson--Lichnerowicz cohomology of a certain target space Poisson structure,
which is isomorphic to the target space de Rham cohomology in the particular 
case of the $A$ model, but it is not in the general case. 

For these reasons, the biKaehler sigma 
model introduced in the present paper is not seemingly related to the 
$(2,2)$ supersymmetric sigma model by twisting in general, at least in the
form defined in \cite{Kapustin2}. This limits 
its relevance for string theory. However, its very existence is
interesting enough from a field theoretic point of view, not least as an 
exemplification of the methodology of ref. \cite{AKSZ}. 

As is well known, any topological field theory (of cohomological type) 
describes the intersection theory of a certain moduli space in terms of 
local quantum field theory. Though we have identified a set of equations, 
which, based on general arguments of topological field theory,  
should define the moduli space underlying the biKaehler sigma model, we 
have no geometrical interpretation and no analytic control on it in general.
It is possible that the biKaehler sigma model found in this paper, 
though satisfying a number of basic formal prerequisites for a consistent
topological field theory, may not pass a closer inspection at the end.  
The investigation of this matter is left for future work. 

The paper is organized as follows. In sect. \ref{sec:biKaehler}, we review 
basic results of biKaehler geometry. In sect. \ref{sec:sigma}, we
introduce the biKaehler sigma model and present its field content 
and its action. In sect. \ref{sec:symmetries}, we
analyze the symmetries of the model and show that it possesses an odd
symmetry $\delta$, that is nilpotent on shell.  In sect. \ref{sec:exactenss},
we prove the topological nature of the model by showing 
that the action is $\delta$--exact on shell up to a topological term, 
when the target space geometry satisfies certain weak restrictions. We further
identify the set of equations describing the underlying moduli space. 
In sect. \ref{sec:cohomology}, we study the local cohomology of $\delta$ 
and show its relation to Poisson--Lichnerowicz cohomology.
In sect. \ref{sec:special}, we write down the action, the symmetries 
and the moduli space equations of the biKaehler sigma model of 
standard biKaehler target geometries and show
that the biKaehler model contains Witten's $A$ model as a particular case.
In sect. \ref{sec:Hitchin}, we show that the biKaehler model 
is obtainable by gauge fixing the Hitchin model with generalized Kaehler 
target space and use this result to show that the associated field theory
depends only on a certain combination of the target space geometrical data. 
Finally, in sect. \ref{sec:discussion} we compare our results with the elegant
geometrical constructions of refs. \cite{Kapustin1, Kapustin2} and 
discuss briefly the open problems. 
\vfill\eject

\begin{small}
\section{\bf BiKaehler geometry}
\label{sec:biKaehler}
\end{small}

Let $M$ be a smooth manifold. An almost biKaehler structure on $M$ consists of
a Riemannian metric $g_{ab}$ and two almost complex structures $K_\pm{}^a{}_b$,
such that $g_{ab}$ is Hermitian with respect to both $K_\pm{}^a{}_b$:
\begin{align}
&K_\pm{}^a{}_cK_\pm{}^c{}_b=-\delta^a{}_b,\vphantom{\int}
\label{K2=-1}
\\
&K_{\pm ab}+K_{\pm ba}=0.\vphantom{\int}
\label{K+Kt=0}
\end{align}
Here and below, indices are raised and lowered by using the metric
$g_{ab}$.
An almost biKaehler structure on $M$ is a biKaehler structure on $M$ 
if the tensors $K_\pm{}^a{}_b$ are parallel with respect to the
Levi--Civita connection $\nabla_a$ of $g_{ab}$ 
\begin{equation}
\nabla_aK_\pm{}^b{}_c=0.
\label{nablaK=0}
\end{equation}
As is well known, this implies that 
the almost complex structures $K_\pm{}^a{}_b$ are integrable and, thus, that 
they are complex structures, and that the metric $g_{ab}$ is Kaehler
with respect to both the $K_\pm{}^a{}_b$.
In the following, we consider only  biKaehler structures. 

The complex structures $K_\pm{}^a{}_b$ can be multiplied, being 
endomorphisms of the tangent bundle $TM$ of $M$. In this way, they 
generate an algebra of endomorphisms $\mathcal{A}$. 
The conventionally normalized anticommutator of the $K_\pm{}^a{}_b$ 
\begin{equation}
C^a{}_b=\frac{1}{2}(K_+{}^a{}_cK_-{}^c{}_b+K_-{}^a{}_cK_+{}^c{}_b)
\label{CC}
\end{equation}
belongs to the center of $\mathcal{A}$. 
By this fact, it is easy to see that the most general element of the algebra
$\mathcal{A}$ is of the form 
\begin{equation}
X^a{}_b=Z^a{}_b+\sum_{\alpha=\pm,0}U_\alpha{}^a{}_cK_\alpha{}^c{}_b
\label{generalX}
\end{equation}
where $K_0{}^a{}_b$ is the conventionally normalized commutator 
of the $K_\pm{}^a{}_b$ 
\begin{equation}
K_0{}^a{}_b=\frac{1}{2}(K_+{}^a{}_cK_-{}^c{}_b-K_-{}^a{}_cK_+{}^c{}_b)
\label{K0}
\end{equation}
and $Z^a{}_b$ and $U_\alpha{}^a{}_b$, $\alpha=\pm,0$, are polynomials in $C^a{}_b$.

By (\ref{K+Kt=0}), the $K_{\pm ab}$ are $2$--forms. By (\ref{nablaK=0}), they 
are parallel and, thus, also closed
\begin{equation}
\partial_{[a}K_{\pm bc]}=0
\label{dKpm=0}
\end{equation}
\footnote{\vphantom{$\bigg[$} Here and below, the brackets $[\cdots]$ denote 
full antisymmetrization of all enclosed 
tensor indices except perhaps for those between bars $|\cdots|$.}. 
They are in fact the Kaehler forms of $g_{ab}$
corresponding to the complex structures $K_\pm{}^a{}_b$. 
Besides the $K_{\pm ab}$, there is another relevant 
$2$--form in biKaehler geometry, $K_{0ab}$.
$K_{0ab}$ is also parallel and, thus, closed
\begin{equation}
\partial_{[a}K_{0bc]}=0.
\label{dK0=0}
\end{equation}
It is not difficult to show that $K_{0ab}$ is of type $(2,0)+(0,2)$ and
holomorphic with respect to both complex structures $K_\pm{}^a{}_b$.

Usually, in Kaehler geometry, it is convenient to write the relevant
tensor identities in the complex coordinates of the underlying complex
structure rather than in general coordinates. 
In biKaehler geometry, one is dealing with two generally non
commuting complex structures. One could similarly write the 
tensor identities in the complex coordinates of either complex structures.
In this case, however, the convenience of complex versus general coordinates
would be limited. We decided, therefore, to opt for general coordinates 
throughout the paper. To this end, we define the complex tensors
\begin{equation}
\Lambda_{\pm}{}^a{}_b=\frac{1}{2}\big(\delta^a{}_b-iK_\pm{}^a{}_b\big).
\label{Pipm}
\end{equation}
The $\Lambda_{\pm}{}^a{}_b$ satisfy the relations 
\begin{subequations}\label{Pipm1}
\begin{align}
&\Lambda_{\pm}{}^a{}_c\Lambda_{\pm}{}^c{}_b=\Lambda_{\pm}{}^a{}_b,\vphantom{\int}
\label{Pipm1a}
\\
&\Lambda_{\pm}{}^a{}_b+\overline{\Lambda}{}_{\pm}{}^a{}_b=\delta^a{}_b,\vphantom{\int}
\label{Pipm1b}
\\
&\Lambda_{\pm}{}^a{}_b=\overline{\Lambda}{}_{\pm}{}_b{}^a.\vphantom{\int}
\label{Pipm1c}
\end{align}
\end{subequations}
Thus, $\Lambda_{\pm}{}^a{}_b$ are projector valued endomorphisms of the
complexified tangent bundle $T_cM$.
The corresponding projection subbundle of $T_cM$ is the $\pm$ holomorphic
tangent bundles $T_\pm^{1,0}M$.
 
The covariant constancy of the complex structures $K_\pm{}^a{}_b$ entails
strong restrictions on the Riemann tensor of the Levi--Civita connection, 
\begin{subequations}\label{Riemann}
\begin{align}
&R_{abce}\Lambda_{\pm}{}^e{}_d=R_{abed}\overline{\Lambda}{}_{\pm}{}^e{}_c,
\vphantom{\int}
\label{Riemanna}
\\
&R_{aecf}\Lambda_{\pm}{}^e{}_{[b}\Lambda_{\pm}{}^f{}_{d]}=0,
\vphantom{\int}
\label{Riemannb}
\\
&\nabla_fR_{abcg}\Lambda_{\pm}{}^f{}_{[d}\Lambda_{\pm}{}^g{}_{e]}=0
\vphantom{\int}
\label{Riemannc}
\end{align}
\end{subequations}
and many other relations following either by complex conjugation or 
from the known symmetry properties of the Riemann tensor.

There are many interesting examples of biKaehler geometries,  which
will be considered in this paper.
A biKaehler structure $g_{ab}$, $K_\pm{}^a{}_b$  
satisfying either conditions 
\begin{subequations}\label{K+=pmK-}
\begin{align}
&K_+{}^a{}_b=-K_-{}^a{}_b, \qquad (K)
\label{K+=-K-}
\\
&K_+{}^a{}_b=K_-{}^a{}_b \,~~\,\qquad (K')
\label{K+=+K-}
\end{align}
\end{subequations}
is obviously equivalent to an ordinary Kaehler structure
$g_{ab}$, $K^a{}_b$, where
\begin{equation}\label{K=K-}
K^a{}_b=K_-{}^a{}_b.
\end{equation}
Thus, there are two ways a Kaehler structure
can be embedded into a biKaehler structure. The resulting biKaehler structures
will be called of type $K$, $K'$ in the following. Conversely, 
a biKaehler structure $g_{ab}$, $K_\pm{}^a{}_b$  
can be viewed as a pair of Kaehler structures with the same underlying metric.

More generally, one can consider biKaehler structures 
$g_{ab}$, $K_\pm{}^a{}_b$ such that
\begin{equation}
K_0{}^a{}_b=0 \qquad(P)
\label{K0=0}
\end{equation}
(cf. eq. (\ref{K0})). (\ref{K0=0}) is equivalent to the statement that
the endomorphisms $K_\pm{}^a{}_b$ commute. We shall call these
biKaehler structures of type $P$. For these
\begin{equation}
L{}^a{}_b=K_+{}^a{}_cK_-{}^c{}_b
\label{prod}
\end{equation}
is a Riemannian product structure of the manifold $M$. The manifold $M$ then factorizes 
locally as a product $M_{+1}\times M_{-1}$ such that the tangent 
bundles $TM_{\pm 1}$ are the $\pm 1$ eigenbundles of the 
endomorphism $L{}^a{}_b$. 
From (\ref{K+=pmK-}), it appears that type $K$, $K'$ biKaehler structures 
are particular cases of type $P$ biKaehler structures. The corresponding
product structures $L^a{}_b=\pm\delta^a{}_b$ are trivial. 

Another important class of biKaehler structures $g_{ab}$, $K_\pm{}^a{}_b$ 
is defined by the condition 
\begin{equation}
C^a{}_b=0.\qquad (HK)
\label{HK}
\end{equation}
(cf. eq. (\ref{CC})), which we shall call of type $HK$. 
(\ref{HK}) is equivalent to the statement that
the endomorphisms $K_\pm{}^a{}_b$ anticommute.
For these, the endomorphisms
\begin{equation}
K_1{}^a{}_b=K_+{}^a{}_b, \quad K_2{}^a{}_b=K_-{}^a{}_b, 
\quad K_3{}^a{}_b=K_0{}^a{}_b
\label{HKstructure}
\end{equation}
form a hyperKaehler structure $g_{ab}$, $K_i{}^a{}_b$, $i=1,~2,~3$.
The manifold $M$ then admits a triplet of Kaehler structures   
with the same underlying metric satisfying the quaternion algebra
\begin{equation}
K_i{}^a{}_cK_j{}^c{}_b=-\delta_{ij}\delta^a{}_b+\epsilon_{ijk}K_k{}^a{}_b.
\label{HKquaternion}
\end{equation}

\vfill\eject

\begin{small}
\section{\bf The biKaehler sigma model}
\label{sec:sigma}
\end{small}

The biKaehler sigma model is a field theoretic realization of biKaehler
geometry.
It is a $2$--dimensional sigma model whose target space
is a manifold $M$ equipped with a biKaehler structure $g_{ab}$,
$K_\pm{}^a{}_b$ and whose world sheet is a Riemann surface $\Sigma$, 
a surface endowed with a complex structure. The fields of the model
are the usual embedding field $x^a$ and three further tensor valued form fields
$y_a$, $\psi^a$, $\chi_a$. They are characterized by their target space and
world sheet global properties and by their ghost degree as summarized by 
the following table.
\begin{equation}\label{fields}
\begin{array}{cccc}
\text{field}\hphantom{xxx}\vphantom{\int}
&\text{global type}&\text{ghost degree} &\text{total degree}\\
x^a \hphantom{xxx}\vphantom{\int}        
&\mathrm{Fun}(\Sigma,M) \hphantom{xxxxx\,\,\,\,}        &0           &0\\
y_a \hphantom{xxx}\vphantom{\int}        
&\Omega^{0,0}(\Sigma,x^*\Pi T^*M) \hphantom{x}          &1           &1\\ 
\psi^a \hphantom{xxx}\vphantom{\int}     
&\Omega^{1,0}(\Sigma,x^*\Pi T_-^{0,1}M)                 &-1          &0\\ 
\chi_a\hphantom{xxx} \vphantom{\int}     
&\Omega^{1,0}(\Sigma,x^*T_-^{*0,1}M) \hphantom{\,\,}    &0           &1
\end{array}
\end{equation}
Here, $\Pi$ is the parity reversion operator of vector bundles, which replaces 
the typical vector fiber with its counterpart of opposite Grassmannality. 
The total degree of a field is the sum of its world sheet form and ghost
degrees and determines its statistics. 
The fields $x^a$, $y_a$ are real. The fields $\psi^a$, $\chi_a$, conversely,
are complex. They are conveniently viewed as elements of 
$\Omega^{1,0}(\Sigma,x^*\Pi T_cM)$, $\Omega^{1,0}(\Sigma,x^*T^*{}_cM)$
satisfying the constraints 
\begin{subequations}\label{constrpsichi}
\begin{align}
\psi^a&=\overline{\Lambda}{}_-{}^a{}_b(x)\psi^b,\vphantom{\int}
\label{constrpsi}
\\
\chi_a&=\Lambda_-{}_a{}^b(x)\chi_b.\vphantom{\int}
\label{constrchi}
\end{align}
\end{subequations}
Note that these constraints break the symmetry of the target space 
biKaehler geometry with respect to the exchange of the two complex structures
$K_\pm{}^a{}_b$. 
Further, they couple the complex structure $K_-{}^a{}_b$ of $M$, 
and the complex structure of $\Sigma$, since the target space global
properties of the fields involved depend on the latter in an essential way. 
We shall analyze these issues in greater detail below in sects. 
\ref{sec:Hitchin}, \ref{sec:discussion}.

The action $S$ of the biKaehler sigma model is given by 
\begin{align}
S&=\int_\Sigma\Big\{-(iP^2\overline{\Lambda}{}_{-ab}+K_{-ab})(x)\barpartial x^a\partial x^b
-ig^{ab}(x)\overline{\chi}{}_a\chi_b
\label{action}
\\
&\hphantom{=\int_\Sigma\Big\{}
+(1+P)J^{ab}(x)
\big(\overline{\chi}{}_a+g_{ac}(x)\barpartial x^c\big)
\big(\chi_b+g_{bd}(x)\partial x^d\big)\vphantom{\int}
\nonumber
\\
&\hphantom{=\int_\Sigma\Big\{}
+M^a{}_b(x)(\psi^b\barnabla y_a+\overline{\psi}{}^b\nabla y_a)
+R^a{}_{ced}P^{eb}(x)\overline{\psi}{}^c\psi^dy_ay_b\Big\},\vphantom{\int}
\nonumber
\end{align}
\footnote{\vphantom{$\bigg[$} Here and below, for 
any number of mixed rank $2$ tensors $A_1{}^a{}_b$, ..., $A_p{}^a{}_b$, we set
$(A_1+A_2+\cdots +A_p)^a{}_b=A_1{}^a{}_b+A_2{}^a{}_b+\cdots 
A_p{}^a{}_b$ and 
$A_1A_2\cdots A_p{}^a{}_b=A_1{}^a{}_{c_1}A_2{}^{c_1}{}_{c_2}\cdots 
A_p{}^{c_{p-1}}{}_b$. Note laso that $(1)^a{}_b=\delta^a{}_b$,
$(1)_{ab}=g_{ab}$, etc.}
where the tensors $J^a{}_b$, $P^a{}_b$, $M^a{}_b$ are given by  
\begin{subequations}
\label{JPM}
\begin{align}
J^a{}_b&=\frac{1}{2}(K_++K_-)^a{}_b, \qquad P^a{}_b=\frac{1}{2}(K_+-K_-)^a{}_b, 
\label{JP}
\\
M^a{}_b&=\frac{1}{2}(1+K_+)(1+K_-)^a{}_b
\label{M}
\end{align}
\end{subequations}
and $\nabla$ is the pull--back by $x^a$ of the Levi--Civita connection 
\begin{equation}\label{nabla}
\nabla=\partial\pm\Gamma^{\mathbf{\cdot}}{}_{\mathbf{\cdot}a}(x)\partial x^a.
\end{equation}
Wedge multiplication of forms is understood.
Due the large number of relations satisfied by the basic tensors of biKaehler
geometry and because of the constraints (\ref{constrpsichi}), $S$ can be
cast in several other equivalent forms. The one shown above is the most
compact, which we were able to find.

The classical field equations associated with the action $S$ are 
easily derived:
\begin{subequations}
\label{fieldeqs}
\begin{align}
&~~~iP^2{}_{ab}(x)\barnabla\partial x^b
+\nabla_aR^b{}_{dfe}P^{fc}(x)\overline{\psi}{}^d\psi^ey_by_c\vphantom{\int}
\label{fieldeqx}
\\
&+\Big[R^d{}_{eca}M^e{}_b(x)\overline{\psi}{}^b\partial x^c
+(1+P)J^b{}_a(x)\barnabla\chi_b+\mathrm{c.c.}\Big]=0,\vphantom{\int}
\nonumber
\\
&~~M^a{}_b(x)\barnabla\psi^b
+R^{[a}{}_{dec}P^{|e|b]}(x)\overline{\psi}{}^c\psi^dy_b+\mathrm{c.c.}=0,\vphantom{\int}
\label{fieldeqpsi}
\\
&~~M\overline{\Lambda}{}_-{}^b{}_a(x)\barnabla y_b
+R^b{}_{fed}\overline{\Lambda}{}_-{}^f{}_aP^{ec}(x)\overline{\psi}{}^dy_by_c=0,
\vphantom{\int}
\label{fieldeqp}
\\
&~~\overline{\Lambda}{}_-(1+P)J^a{}_b(x)\barpartial x^b
+\overline{\Lambda}{}_-P(1+J)^{ab}(x)\overline{\chi}{}_b=0.\vphantom{\int}
\label{fieldeqchi}
\end{align}
\end{subequations}
In obtaining (\ref{fieldeqp}), (\ref{fieldeqchi}), one must take
into due account the constraints (\ref{constrpsichi}).

The action $S$ may be modified by the addition of topological terms of the
form \hphantom{xxxxxxxxxxxxxxxxxxxxxxxxx}
\begin{equation}
S_{\mathrm{top}}=\int_\Sigma x^*\omega
=\int_\Sigma \omega_{ab}(x)\barpartial x^a\partial x^b,
\label{actiontop}
\end{equation}
where $\omega_{ab}$ is a closed $2$--form, without changing the field
equations and the infinitesimal symmetries of the action. For instance
\begin{equation}
\omega_{ab}=\sum_{\alpha=\pm,0} c_\alpha K_{\alpha ab},
\label{omegaexample}
\end{equation}
where the $c_\alpha$ are real coefficients. The terms 
\begin{equation}
\int_\Sigma\big(-K_-+(1+P)J\big){}_{ab}(x)\barpartial x^a\partial x^b
\end{equation}
appearing in the expression of the action $S$, eq. (\ref{action}), are
precisely of this form. Thus their inclusion is somewhat conventional at this stage.
\vfill\eject

\begin{small}
\section{\bf The symmetries of the model}
\label{sec:symmetries}
\end{small}

The biKaehler sigma model action $S$ introduced in sect. \ref{sec:sigma} 
exhibits a bosonic symmetry associated with the following infinitesimal 
even variations
\begin{subequations}
\label{delta'fields}
\begin{align}
\delta_{\mathrm{gh}}x^a&=0,\vphantom{\int}
\label{delta'x}
\\
\delta_{\mathrm{gh}}y_a&=-y_a,\vphantom{\int}
\label{delta'p}
\\
\delta_{\mathrm{gh}}\psi^a&=\psi^a,\vphantom{\int}
\label{delta'psi}
\\
\delta_{\mathrm{gh}}\chi_a&=0,\vphantom{\int}
\label{delta'chi}
\end{align}
\end{subequations}
where multiplication by an infinitesimal real even parameter is tacitly understood, so
that \hphantom{xxxxxxxxxxxxxxxxxxxxxxxxxxxxx}
\begin{equation}
\delta_{\mathrm{gh}}S=0.
\label{delta'S=0}
\end{equation}
It is easy to see that this nothing but ghost number symmetry.
The associated symmetry current is \hphantom{xxxxxxxxxxxxxxxxxxx}
\begin{equation}
\mathcal{I}=-iM^a{}_b(x)\psi^by_a+\mathrm{c.c.}
\label{delta'current}
\end{equation}
as is easily verified.

The action $S$ exhibits also a fermionic symmetry associated with the 
following infinitesimal odd variations
\begin{subequations}
\label{deltafields}
\begin{align}
\delta x^a&=P^{ab}(x)y_b,\vphantom{\int}
\label{deltax}
\\
\delta y_a&=-\Gamma^b{}_{ad}P^{dc}(x)y_by_c,\vphantom{\int}
\label{deltay}
\\
\delta\psi^a&=-\Gamma^a{}_{bd}P^{dc}(x)\psi^by_c
+\overline{\Lambda}{}_-(J^2+J+1)^a{}_b(x)\partial x^b
+\overline{\Lambda}{}_-P^{ab}(x)\chi_b,\vphantom{\int}
\label{deltapsi}
\\
\delta\chi_a&=-\Gamma^b{}_{ad}P^{dc}(x)\chi_by_c
-\Lambda_-J_a{}^b(x)\nabla y_b
+\Lambda_{-a}{}^fR^b{}_{fed}P^{ec}(x)\psi^dy_by_c,\vphantom{\int}
\label{deltachi}
\end{align}
\end{subequations}
where multiplication by an infinitesimal real odd parameter is tacitly understood, 
so that \hphantom{xxxxxxxxxxxxxxxxxxxxxxxxxxxxx}
\begin{equation}
\delta S=0.
\label{deltaS=0}
\end{equation}
The verification of (\ref{deltaS=0}) is lengthy but totally straightforward.
The associated symmetry current is
\begin{equation}
\mathcal{S}=\big(i(J^2+J+1)J-P^2P\overline{\Lambda}{}_-\big){}^a{}_b(x)\partial x^bp_a
+i(P^2+P)J^{ab}(x)\chi_ap_b+\mathrm{c.c.}
\label{deltacurrent}
\end{equation}

The symmetry $\delta$ is nilpotent on shell, as it appears from the following 
computation
\begin{subequations}
\label{delta2fields}
\begin{align}
\delta^2 x^a&=0,\vphantom{\int}
\label{delta2x}
\\
\delta^2 y_a&=0,\vphantom{\int}
\label{delta2p}
\\
\delta^2\psi^a&=-i\overline{\Lambda}{}_+\Lambda_-{}^{ga}(x)\Big[
M\Lambda{}_-{}^b{}_g(x)\nabla y_b
+R^b{}_{fed}\Lambda_-{}^f{}_gP^{ec}(x)\psi^dy_by_c\Big],\vphantom{\int}
\label{delta2psi}
\\
\delta^2\chi_a&=R^c{}_{feh}\overline{\Lambda}{}_-{}^f{}_aP^{hd}(x)
\Big[\Lambda_-(1+P)J^e{}_b(x)\partial x^b\vphantom{\int}
\label{delta2chi}
\\
&\hskip6cm+\Lambda_-P(1+J)^{eb}(x)\chi_b\Big]y_cy_d\vphantom{\int}
\nonumber
\end{align}
\end{subequations}
and from (\ref{fieldeqp}), (\ref{fieldeqchi}).
The verification of (\ref{delta2fields}) is also lengthy but straightforward.
We note that only two of the four field equations, namely (\ref{fieldeqp}), (\ref{fieldeqchi}),
are involved. In more precise terms, (\ref{delta2fields}) states that $\delta$
is nilpotent on the quotient of the algebra of all field functionals by 
the bilateral ideal generated by the field equations (\ref{fieldeqp}), (\ref{fieldeqchi}).
If we denote by $\approx$ equality on the quotient algebra, we may write
\hphantom{xxxxxxxxxxxxxxxxxxxxxx}
\begin{equation}\label{delta2=0}
\delta^2\approx 0. 
\end{equation}
However, for the sake of brevity, one says simply that $\delta$ is nilpotent on shell. 
The study the cohomology associated with $\delta$ is naturally the next step of
our analysis.

\vfill\eject

\begin{small}
\section{\bf The topological nature of the model}
\label{sec:exactenss}
\end{small}

We have seen above that the biKaehler sigma model is a sigma model with 
an odd symmetry that is nilpotent on shell
(cf. eqs. (\ref{deltaS=0}), (\ref{delta2=0})). 
This makes it akin to some 
extent to the existent topological models \cite{Witten1,Witten2}. 
These latter however have a further property, that is crucial 
to ensure their topological nature: the action is $\delta$ exact
on shell up to topological terms.
The natural question arises about whether the biKaehler sigma model 
we illustrated above has the same property.

A gauge fermion $\Psi$ is a local functional of the fields of ghost number $-1$.
We are looking for a gauge fermion $\Psi$ such that
\begin{equation}
S\approx\delta\Psi+S_\mathrm{top}, 
\label{gaugefermiondef}
\end{equation}
where $S_\mathrm{top}$ is some topological functional of $x^a$
of the form (\ref{actiontop}) and $\approx$ denotes equality on shell
in the sense explained at the end of sect. \ref{sec:symmetries}.
A gauge fermion $\Psi$ with the above property exists
for the $A$ topological sigma model, as shown by Witten long ago
\cite{Witten1,Witten2}. It is natural to wonder whether
a gauge fermion $\Psi$ exists for the biKaehler sigma model.
We found this problem unexpectedly difficult. In fact, we have not been able
to show that such a $\Psi$ exists for an arbitrary biKaehler target
geometry. However, as we show below, we succeeded in finding a $\Psi$
such that \hphantom{XXXXXXXXXXXXXXXX}
\begin{equation}
S\approx\delta\Psi+S_\mathrm{top}+\Omega, 
\label{gaugefermionanomaly}
\end{equation}
where $\Omega$ is a ``topological anomaly'', which is generally non vanishing,
but which does vanish for a subclass of biKaehler structures, defined by a week 
condition, which moreover contains all the standard examples illustrated in
sect. \ref{sec:biKaehler}. 

Two scenarios are thus possible. In the first scenario, a gauge fermion $\Psi$
satisfying (\ref{gaugefermiondef}) exists, but it is rather complicated,
making its computation prohibitively difficult. In the second scenario, a 
gauge fermion $\Psi$ does not exist in general. In such a case, it would 
be important to characterize the corresponding biKaehler structures.

To begin with, we recall that we are tackling a cohomological problem, 
so that its solution, if it exists, is certainly not unique, but it is
affected by the customary cohomological ambiguities. 
We may start with an ansatz of the form
\begin{equation}
\Psi=\int_\Sigma\frac{i}{2}\Big[
A_{ab}(x)(\overline{\psi}{}^a\partial x^b-\psi^a\barpartial x^b)
+B^a{}_b(x)(\overline{\chi}_a\psi^b-\chi_a\overline{\psi}{}^b)\Big],
\label{gaugefermion}
\end{equation}
where $A_{ab}$, $B^a{}_b$ are real tensors satisfying
\begin{equation}
\nabla_cA_{ab}=0,\qquad \nabla_cB^a{}_b=0.
\label{nablaAB=0}
\end{equation}
Next, we compute $\delta\Psi$ using (\ref{deltafields})
and simplify the resulting expression using
the constraints (\ref{constrpsichi}) and the field equations (\ref{fieldeqp}),
(\ref{fieldeqchi}) only. Finally, by a procedure of trial and error, we 
adjust the expressions of $A_{ab}$, $B^a{}_b$, in such a way to enforce  
(\ref{gaugefermiondef}) or (\ref{gaugefermionanomaly}).
In this way, we find that a relation of the form (\ref{gaugefermionanomaly})
holds if $A_{ab}$, $B^a{}_b$ are given by 
\begin{subequations}\label{AB}
\begin{align}
A_{ab}&=g_{ab}+\frac{1}{Z^2-16}(-1-C+4P)J(1+C+4J-4P)_{ab},\label{A}\\
B^a{}_b&=(1+P)J^a{}_b-\frac{4}{Z^2-16}(1+C)(2+P)J^a{}_b\label{B},
\end{align}
\end{subequations}
where \hphantom{XXXXXXXXXXXXXX}
\begin{equation}
C{}^a{}_b=(J^2-P^2)^a{}_b
\label{C=J2-P2}
\end{equation} 
is nothing but the central tensor (\ref{CC}) and $Z{}^a{}_b$ is some function 
of $C{}^a{}_b$ subject to the only condition that the endomorphisms
$(Z\pm 4){}^a{}_b$ are
pointwise invertible on $M$. In such a case, the topological term
$S_\mathrm{top}$ is given by 
\begin{align}
S_\mathrm{top}&=\int_\Sigma\Big[\frac{1}{2}P(2+J)_{ab}(x)
+\frac{1}{2}\Big(\frac{\pm 1}{Z\pm 4}-\frac{1}{Z^2-16}(C\pm Z+5)\Big)\label{Stoptop}\\
&\hskip3cm \times P\big(2(-1+C)+(1+C)J\big){}_{ab}(x)
\Big]\barpartial x^a\partial x^b,\vphantom{\int}\nonumber
\end{align}
while the topological anomaly $\Omega$ reads 
\begin{align}
\Omega&=\int_\Sigma\frac{i}{2(Z^2-16)}\big((C+5)^2-Z^2\big)PJ^a{}_b(x)\label{Omegaexpl}\\
&\hskip2.5cm \times\big(\chi_a\barpartial x^b-\overline{\chi}{}_a\partial x^b
+\overline{\psi}{}^b\nabla y_a-\psi^b\barnabla y_a\big).\vphantom{\int}\nonumber
\end{align}
We note that the integrand of $S_\mathrm{top}$ is indeed a closed form, since
the tensors $J_{ab}$, $P_{ab}$, $PJ_{ab}$ are antisymmetric, the tensors 
$C_{ab}$, $Z_{ab}$ are symmetric and central and all are covariantly
constant. Furthermore, $S_\mathrm{top}$ does not depend on the sign choice 
in the integrand, as is easy to check.

The topological anomaly $\Omega$ vanishes in a number of cases. If $(C+5\pm 4)^a{}_b$
is pointwise invertible on $M$, we can chose 
\begin{equation}
Z^a{}_b=\pm(C+5)^a{}_b
\label{Z=C+5}
\end{equation} 
and make $\Omega$ vanish. The biKaehler structures of type $HK$
(see sect. \ref{sec:biKaehler}) fall in this category, since,
for these, $C^a{}_b=0$. Alternatively, we see that $\Omega$ vanishes 
when $PJ^a{}_b=0$. The biKaehler structures of type $P$, in particular 
those of types $K$, $K'$, (see sect. \ref{sec:biKaehler}),  have this property, 
since 
\begin{equation}
K_0{}^a{}_b=2PJ^a{}_b
\label{PJ=K0}
\end{equation} 
and, for these, $K_0{}^a{}_b=0$. 
It would be interesting to characterize the biKaehler structures, if any,
for which the topological anomaly $\Omega$ fails to vanish. 

When the target space biKaehler geometry is such that the topological 
anomaly $\Omega$ does indeed vanish, we expect the corresponding biKaehler 
sigma model to be a topological field theory, in analogy 
to what happens in Witten's $A$ sigma model \cite{Witten1,Witten2}. 
In the $A$ model, the topological correlators are independent 
from the world sheet complex structure and from the target manifold 
complex structure, but they do depend on the target manifold 
symplectic structure. For similar reasons, one would expect
the biKaehler model topological correlators to be independent 
from the world sheet complex structure and to depend only on a proper 
subset of the target space geometrical data. 
In sect. \ref{sec:Hitchin}, we shall identify precisely this latter. 

The above analysis provides strong evidence that the biKaehler sigma model 
might indeed be a topological field theory akin to the $A$ sigma model.
One of the most basic features of topological field theories of cohomological
type is that the functional measure of the associated quantum field theories 
localizes on the space of field configurations which are fixed point for the 
topological BRST charge \cite{Witten3}. In our case, these are the field 
configurations $x^a$, $y_a$, $\psi^a$, $\chi_a$ satisfying 
\begin{subequations}
\label{deltafields=0}
\begin{align}
\delta x^a&=0,\vphantom{\int}
\label{deltax=0}
\\
\delta y_a&=0,\vphantom{\int}
\label{deltay=0}
\\
\delta\psi^a&=0,\vphantom{\int}
\label{deltapsi=0}
\\
\delta\chi_a&=0.\vphantom{\int}
\label{deltachi=0}
\end{align}
\end{subequations}
From (\ref{deltafields}), the (\ref{deltafields=0}) are equivalent
to the following set of equations
\begin{subequations}
\label{moduli}
\begin{align}
&P^{ab}(x)y_b=0,\vphantom{\int}
\label{moduli1}
\\
&\Lambda_-(J^2+J+1)^a{}_b(x)\barpartial x^b
+\Lambda_-P^{ab}(x)\overline{\chi}{}_b=0,\vphantom{\int}
\label{moduli2}
\\
&\overline{\Lambda}{}_-J_a{}^b(x)\barnabla y_b=0.\vphantom{\int}
\label{moduli3}
\end{align}
\end{subequations}
The geometrical interpretation of these equations is not known to us, except
for certain particular cases. We expect also that they may suffer some kind 
of disease for the biKaehler structures, for which the topological anomaly 
$\Omega$ does not vanish (if any), and, perhaps, for an even larger class of 
such structures. A detailed investigation of these matters is
beyond the scope of this paper. Here, we shall restrict ourselves to making a few 
general observations. When the endomorphism $P^a{}_b$ is pointwise invertible on $M$
(e. g. for a type $K$ or $HK$ biKaehler structure), eq. (\ref{moduli1})
becomes equivalent to the equation
\begin{equation}
y_a=0
\label{y=0}
\end{equation} 
and eq. (\ref{moduli3}) is identically satisfied.
Eq. (\ref{moduli2}) is a kind of generalized holomorphy condition for the
embedding field $x^a$.  For a type $P$  biKaehler structure, it reduces to 
\begin{equation}
\Lambda_-(J^2+J+1)^a{}_b(x)\barpartial x^b=0,
\label{JJdbarx=0}
\end{equation} 
on account of (\ref{constrchi}). In particular, for a type $K$, $K'$ 
structure, it yields a the customary notion of holomorphy
\begin{equation}
\Lambda_-{}^a{}_b(x)\barpartial x^b=0.
\label{dbarx=0}
\end{equation} 

We conclude this section by recalling that, in the $A$ model, the 
field configurations annihilated by the topological BRST charge
make the gauge fermion $\Psi$ vanish. Apparently, a similar property 
does not hold in general for the gauge fermion $\Psi$ of the biKaehler 
sigma model found above. We do not know whether this is a cohomological
artifact of such $\Psi$ or, else, it is a basic feature of the model. 

\vfill\eject

\begin{small}
\section{\bf The local cohomology of $\delta$}
\label{sec:cohomology}
\end{small}

In this section, we shall study some aspects of the local cohomology of $\delta$.
In view of the topological nature of the biKaehler sigma model, shown 
above, this is an important step of our analysis, because of its relevance for
the classification of topological observables and the study of the properties
of their correlators.

Relations (\ref{deltax}), (\ref{deltay}) and (\ref{delta2x}), (\ref{delta2p}) show that
the fields $x^a$, $y_a$ generate a subcohomology of the $\delta$ cohomology,
which we shall analyze next. For any $p$--vector $X^{a_1\ldots a_p}$, set
\hphantom{xxxxxxxxxxxx}
\begin{equation}\label{Xhat}
\mathcal{O}_{X}
=\frac{1}{p!}X^{a_1\ldots a_p}(x)y_{a_1}\ldots y_{a_p}.
\end{equation}
This is the most general local field containing only the fields 
$x^a$, $y_a$ and no derivatives. Further, it is evident that $\mathcal{O}$ 
maps isomorphically the algebra of multivectors into the algebra of such local
fields formed with $x^a$, $y_a$. Using (\ref{deltax}), (\ref{deltay}) 
and the fact that $\nabla_cP^{ab}=0$, it is easy to show that 
\begin{equation}\label{deltaXhat}
\delta\mathcal{O}_{X}=\mathcal{O}_{\sigma_{\mathrm{PL}}X},
\end{equation}
where $\sigma_{\mathrm{PL}}X^{a_1\ldots a_{p+1}}$ is the $p+1$--vector given by
\begin{equation}\label{sigmaX}
\sigma_{\mathrm{PL}}X^{a_1\ldots a_{p+1}}=
-(p+1)P^{[a_1|c|}\partial_cX^{a_2\ldots a_{p+1}]}
+\frac{(p+1)p}{2}\partial_cP^{[a_1a_2}X^{|c|a_3\ldots a_{p+1}]}.
\end{equation}
Thus, 
\begin{subequations}\label{deltaxy}
\begin{align}
&\delta\mathcal{O}_{X}=0 
~~~\Leftrightarrow~~~\sigma_{\mathrm{PL}}X^{a_1\ldots a_{p+1}}=0, 
\label{deltaXhat=0}\\
&\mathcal{O}_{X}=\delta\mathcal{O}_{Y}
~~~\Leftrightarrow~~~X^{a_1\ldots a_p}=\sigma_{\mathrm{PL}}Y^{a_1\ldots a_p}.
\label{Xhat=deltayhat}
\end{align}
\end{subequations}
The above construction has a simple geometric interpretation. 
Since $\nabla_cP^{ab}=0$, $P^{ab}$ is a Poisson $2$--vector 
on $M$ defining a Poisson structure $P$
\cite{Vaisman}. $\sigma_{\mathrm{PL}}$ is the well known associated 
nilpotent Poisson--Lichnerowicz operator on multivectors \cite{Vaisman}. 
(\ref{deltaXhat}) shows that $\mathcal{O}$ is a cochain isomorphism of
the Poisson--Lichnerowicz multivector cochain complex, 
$({\scri V}^*(M),\sigma_{\mathrm{PL}})$, into the chain complex 
of local fields formed by $x^a$, $y_a$ with no derivatives, $({\scri F}_{xy},\delta)$. 
Thus, the cohomology of the latter, $H^*({\scri F}_{xy},\delta)$, 
is isomorphic to the Poisson--Lichnerowicz multivector cohomology
$H^*_{\mathrm{PL}}(M,P)$.  

For any $p$--form $\omega_{a_1\ldots a_p}$, define the $p$--vector
\begin{equation}\label{sharpomega}
\#\omega^{a_1\ldots a_p}=P^{a_1b_1}\ldots P^{a_pb_p}\omega_{b_1\ldots b_p}.
\end{equation}
$\#$ maps the algebra of forms into the algebra of multivectors. 
As is well known, $\#$ defines a cochain homomorphism of the 
de Rham differential form cochain complex, $(\Omega^*(M),d_{\mathrm{dR}})$, into the 
Poisson--Lichnerowicz multivector cochain complex 
$({\scri V}^*(M),\sigma_{\mathrm{PL}})$, and, thus, also a homomorphism of the 
de Rham cohomology, $H^*_{\mathrm{dR}}(M)$, into the Poisson--Lichnerowicz 
cohomology, $H^*_{\mathrm{PL}}(M,P)$, \cite{Vaisman}. 
This homomorphism is an isomorphism, if $P$ is pointwise invertible, 
i. e. it comes from a symplectic structure. 
Composing the maps $\#$ and $\mathcal{O}$, we have a homomorphism 
of $H^*_{\mathrm{dR}}(M)$ into $H^*({\scri F}_{xy},\delta)$, which is 
an isomorphism when $P$ is pointwise invertible.  

Let $X^{a_1\ldots a_p}$ be a $p$--vector such that $\sigma_{\mathrm{PL}}X^{a_1\ldots a_{p+1}}=0$.
Then, by (\ref{deltaXhat}), $\delta\mathcal{O}_X=0$.
Starting from $\mathcal{O}_X$, one can generate a triplet of
local $\delta$ cohomology classes, by using the well known descent formalism
\cite{Witten1,Witten2}. This is based on the mod $d$ cohomology of $\delta$,
or, equivalently, on the cohomology of $\delta+d$, where $d$ is the de Rham
differential of $\Sigma$. Let us write $\mathcal{O}_X$ as
$\mathcal{O}^{(0)}_X$ to emphasize the fact that it is a $0$--form
on $\Sigma$. We know that \hphantom{xxxxxxxxxxxxxxxxxxxxxx}
\begin{equation}\label{deltaO0=0}
\delta\mathcal{O}^{(0)}_X=0.
\end{equation}
We can integrate $\mathcal{O}^{(0)}_X$ on any $0$--cycle $\Delta$ of
$\Sigma$ (that is evaluate it on a formal sum of points of $\Sigma$), 
yielding an object \hphantom{xxxxxxxxxxxxxxxxxxxxxx}
\begin{equation}\label{O0Delta}
\mathcal{O}^{(0)}_X(\Delta)=\oint_\Delta\mathcal{O}^{(0)}_X.
\end{equation}
By (\ref{deltaO0=0}), $\mathcal{O}^{(0)}_X(\Delta)$ satisfies clearly
\hphantom{xxxxxxxxxxxxxxxxxxxxxx}
\begin{equation}\label{deltaO0Delta=0}
\delta\mathcal{O}^{(0)}_X(\Delta)=0
\end{equation}
and, so, defines a first local $\delta$ cohomology class. This class depends only on
the homology class of the $0$--cycle $\Delta$, if $\mathcal{O}^{(0)}_X(\Delta)$
changes by a local $\delta$ exact term, when $\Delta$ changes by a $0$--boundary.
By Stokes' theorem, this happens provided there exists a local $1$--form 
$\mathcal{O}^{(1)}_X$ field such that \hphantom{xxxxxxxxxxxxxxxxxxxxxx}
\begin{equation}\label{dO0=deltaO1}
d\mathcal{O}^{(0)}_X=\delta\mathcal{O}^{(1)}_X. 
\end{equation}
Let us assume this. We can integrate $\mathcal{O}^{(1)}_X$ on any $1$--cycle 
$\Gamma$ of $\Sigma$ (roughly a formal sum of closed oriented paths
on $\Sigma$), obtaining an object  
\hphantom{xxxxxxxxxxxxxxxxxxxxxx}
\begin{equation}\label{O1Gamma}
\mathcal{O}^{(1)}_X(\Gamma)=\oint_\Gamma\mathcal{O}^{(1)}_X.
\end{equation}
By (\ref{dO0=deltaO1}) and Stokes' theorem, $\mathcal{O}^{(1)}_X(\Gamma)$
satisfies 
\hphantom{xxxxxxxxxxxxxxxx}
\begin{equation}\label{deltaO1Gamma=0}
\delta\mathcal{O}^{(1)}_X(\Gamma)=0
\end{equation}
and, so, defines a second local $\delta$ cohomology class. By Stokes' theorem again, 
this class depends only on the homology class of the $1$--cycle $\Gamma$, 
if $\mathcal{O}^{(1)}_X(\Gamma)$ changes by a local $\delta$ exact term, 
when $\Gamma$ changes by a $1$--boundary. This happens if there is a local 
$2$--form $\mathcal{O}^{(2)}_X$ field such that \hphantom{xxxxxxxxxxxxxxxxxxxxxx}
\begin{equation}\label{dO1=deltaO2}
d\mathcal{O}^{(1)}_X=\delta\mathcal{O}^{(2)}_X. 
\end{equation}
We can integrate $\mathcal{O}^{(2)}_X$ on $\Sigma$, yielding
\hphantom{xxxxxxxxxxxx}
\begin{equation}\label{O1Sigma}
\mathcal{O}^{(2)}_X(\Sigma)=\oint_\Sigma\mathcal{O}^{(2)}_X.
\end{equation}
By (\ref{dO1=deltaO2}) and Stokes' theorem, $\mathcal{O}^{(2)}_X(\Sigma)$ 
satisfies \hphantom{xxxxxxxxxxxxxxxxxxx}
\begin{equation}\label{deltaO2Sigma=0}
\delta\mathcal{O}^{(2)}_X(\Sigma)=0
\end{equation}
and, so, defines a third $\delta$ cohomology class.
Clearly, since $\Sigma$ is $2$--dimensional, the iterative procedure outlined above
stops here. Next, let us find expressions for the descendant fields
$\mathcal{O}^{(q)}_X$, $q=0,1,2$.

As an ansatz, we write
\begin{equation}\label{Xp}
\mathcal{O}^{(q)}_X
=\frac{1}{q!(p-q)!}X^{(q)}{}_{b_1\ldots b_q}{}^{a_1\ldots a_{p-q}}(x)  
dx^{b_1}\ldots dx^{b_q}y_{a_1}\ldots y_{a_{p-q}},
\end{equation}
where the tensor $X^{(q)}{}_{b_1\ldots b_q}{}^{a_1\ldots a_{p-q}}$ is a
$q$--form $p-q$--vector (i. e. it is antisymmetric in the upper and lower indices) and $q=0,1,2$. 
Then, by (\ref{deltax}), (\ref{deltay}), the descent equations (\ref{deltaO0=0}), 
(\ref{dO0=deltaO1}), (\ref{dO1=deltaO2}) hold provided
\begin{equation}\label{X2=PX1}
X^{(q-1)}{}_{b_1\ldots b_{q-1}}{}^{a_1\ldots a_{p-q+1}}
-P^{a_1c}X^{(q)}{}_{b_1\ldots b_{q-1}c}{}^{a_2\ldots a_{p-q+1}}=0,\vphantom{\int}
\end{equation}
for $q=1,2$ and
\begin{equation}\label{sigmaPLXq-1=dXq}
\sigma_{\mathrm{PL}}X^{(q)}{}_{b_1\ldots b_q}{}^{a_1\ldots a_{p-q+1}}
-(-1)^qq\partial_{[b_1}X^{(q-1)}{}_{b_2\ldots b_q]}{}^{a_1\ldots a_{p-q+1}}=0,
\end{equation}
for  $q=0,1,2$, where, for a generic $q$--form $p$--vector 
$X_{b_1\ldots b_q}{}^{a_1\ldots a_p}$
\begin{align}
&\sigma_{\mathrm{PL}}X_{b_1\ldots b_q}{}^{a_1\ldots a_{p+1}}
=-(p+1)P^{[a_1|c|}\partial_cX_{b_1\ldots b_q}{}^{a_2\ldots a_{p+1}]}\vphantom{\int}\\
&+\frac{p(p+1)}{2}\partial_cP^{[a_1a_2}X_{b_1\ldots b_q}{}^{|c|a_3\ldots a_{p+1}]}
+(-1)^q(p+1)q\partial_{[b_1}P^{[a_1|c|}X_{b_2\ldots b_q]c}{}^{a_2\ldots a_{p+1}]}.\vphantom{\int}\nonumber
\end{align}

Above, $\sigma_{\mathrm{PL}}$ is the generalized Poisson--Lichnerowicz operator. 
It acts naturally and is covariantly defined on the space of $q$--form
$p$--vectors $X_{b_1\ldots b_q}{}^{a_1\ldots a_p}$ 
satsifying the algebraic constraint \hphantom{xxxxxxxxxxxxxxxxxxxxxx}
\begin{equation}\label{PX=0}
P^{ab}X_{b_1\ldots b_{q-1}b}{}^{a_1\ldots a_p}=0,
\end{equation}
for $q\geq 1$. It can be shown that $\sigma_{\mathrm{PL}}X_{b_1\ldots b_q}{}^{a_1\ldots a_{p+1}}$
is a $q$--form $p+1$--vector fulfilling (\ref{PX=0}) as well and that 
$\sigma_{\mathrm{PL}}\sigma_{\mathrm{PL}}X_{b_1\ldots b_q}{}^{a_1\ldots a_{p+2}}=0$. 
Thus, if we denote by ${\scri V}_q^p(M)$ the space of $q$--form $p$--vector
tensors for which (\ref{PX=0}) holds, we have a generalized Poisson--Lichnerowicz 
$q$--form multivector cochain complex $({\scri V}_q^*(M),\sigma_{\mathrm{PL}})$, 
and, associated with it, a generalized Poisson--Lichnerowicz 
$q$--form multivector cohomology $H^*_{\mathrm{PL}q}(M,P)$, for every $q\geq 0$. 
The complex and its cohomology are trivial for $q\geq 1$, if $P$ is
pointwise invertible and, so, comes from a symplectic structure, but they are
not so in general. 

In (\ref{sigmaPLXq-1=dXq}), the tensor $X^{(q)}{}_{b_1\ldots b_q}{}^{a_1\ldots a_{p-q}}$
satisfies the constraint (\ref{X2=PX1}) rather than (\ref{PX=0}), and, therefore, 
$\sigma_{\mathrm{PL}}X^{(q)}{}_{b_1\ldots b_q}{}^{a_1\ldots a_{p-q+1}}$ is not covariantly defined. 
However, the combination of the two terms of the left hand side of
(\ref{sigmaPLXq-1=dXq}) is covariant, so that (\ref{sigmaPLXq-1=dXq}) makes sense.

We do not know any general conditions ensuring the existence of solutions of the
descent equations  (\ref{X2=PX1}), (\ref{sigmaPLXq-1=dXq}). However, it is not
difficult to show that there exists a solution when the $p$--vector 
$X^{a_1\ldots a_p}$ we start with is of the form $\#\omega^{a_1\ldots a_p}$
(cf. eq. (\ref{sharpomega})), for some closed $p$--form  $\omega_{a_1\ldots a_p}$,
\begin{equation}\label{domega=0}
\partial_{[a_1}\omega_{a_2\ldots a_{p+1}]}=0.
\end{equation}
Indeed, in this case, the $q$--form $p$--vectors 
\begin{equation}\label{Xq=Pomega}
X^{(q)}{}_{b_1\ldots b_q}{}^{a_1\ldots a_{p-q}}
=P^{a_1c_1}\ldots P^{a_{p-q}c_{p-q}}\omega_{b_1\ldots b_qc_1\ldots c_{p-q}},
\end{equation}
$q=0,1,2$, satisfy eqs. (\ref{X2=PX1}), (\ref{sigmaPLXq-1=dXq}), as is
straightforward to verify. So, when $P^{ab}$ comes from a symplectic structure, we
recover the usual de Rham descent sequence. 

If a $q$--form $p-q$--vector $X^{(q)}{}_{b_1\ldots b_q}{}^{a_1\ldots a_{p-q}}$ 
satisfies (\ref{PX=0}), then the field $\mathcal{O}^{(q)}_X$, given by (\ref{Xp}),
has the property that
\begin{equation}
\delta \mathcal{O}^{(q)}_X=\mathcal{O}^{(q)}_{\sigma_{\mathrm{PL}}X}.
\end{equation}
So, if the generalized Poisson--Lichnerowicz cohomology spaces
$H^{p-q}_{\mathrm{PL}q}(M,P)$, $q=1,2$, do not vanish, the descent 
sequence $\mathcal{O}^{(0)}_X(\Delta)$, $\mathcal{O}^{(1)}_X(\Gamma)$, 
$\mathcal{O}^{(2)}_X(\Sigma)$ constructed above is not uniquely determined by
$X^{a_1\ldots a_p}$ and the cycles $\Delta$, $\Gamma$, $\Sigma$. 
This is, we believe, a novel feature of the biKaehler model.  

We remark that the cohomological setup expounded above depends on the
target space biKaehler geometrical data only through the combination $P^{ab}$. 
This reflects the topological nature of the associated field theory. 

Of course, the above analysis does not exhaust the whole local $\delta$ cohomology. 
A full computation of the cohomology would be interesting, but, unfortunately, 
it is rather difficult because $\delta$ is nilpotent only on shell.

\vfill\eject

\begin{small}
\section{\bf Special biKaehler sigma models}
\label{sec:special}
\end{small}

In this section, we shall consider the biKaehler sigma models associated with the
special biKaehler structures considered at the second half of sect. 
\ref{sec:biKaehler}, since we expect these models to have
special properties, which call for a closer inspection. 
We shall also find that one of these models is just
Witten's $A$ topological sigma model \cite{Witten1,Witten2}. 

We consider first a biKaehler structures $g_{ab}$, $K_\pm{}^a{}_b$ 
of the closely related types $K$, $K'$ and $P$ 
(cf. eqs. (\ref{K+=pmK-}), (\ref{K0=0})).
Recall that a biKaehler structure of type $K$, $K'$ 
corresponds to an ordinary Kaehler structure $g_{ab}$, $K^a{}_b$,
where $K^a{}_b$ is given by eq. (\ref{K=K-}). 
Recall also that a biKaehler structure of type $P$
induces a Riemannian product structure $g_{ab}$, $L^a{}_b$, 
where $L^a{}_b$ is given by eq. (\ref{prod}). 
Finally, recall that a biKaehler structure of type $K$, $K'$ 
is also a particular biKaehler structure of type $P$ for which
$L^a{}_b=\delta^a{}_b,-\delta^a{}_b$, respectively. 
Below, for all these three types of structures, we shall set 
\begin{equation}\label{K=K-again}
K^a{}_b=K_-{}^a{}_b
\end{equation}
and \hphantom{xxxxxxxxxxxxxxxxxxxxxxxxxxxxxxxx}
\begin{equation}\label{Lambda=Lambda-}
\Lambda^a{}_b=\Lambda_-{}^a{}_b
\end{equation}
(cf. eqs. (\ref{K=K-}), (\ref{Pipm})). We further set
\begin{equation}\label{Ppm}
Q_\pm{}^a{}_b=\frac{1}{2}(1\pm L)^a{}_b.
\end{equation}
$Q_\pm{}^a{}_b$ is the orthogonal projector on the $\pm 1$ eigenbundle of 
$L^a{}_b$. $Q_-{}^a{}_b=0$, $Q_+{}^a{}_b=0$, when the type $P$ structure
is a type $K$, $K'$ structure, respectively.

From (\ref{constrpsichi}), the fields $\psi^a$, $\chi_a$ of the associated
biKaehler sigma models satisfy the constraints \hphantom{xxxxxxxxxxxxxxxxxx}
\begin{subequations}\label{KK'Pconstrpsichi}
\begin{align}
\psi^a&=\overline{\Lambda}^a{}_b(x)\psi^b\vphantom{\int},
\label{KK'Pconstrpsi}
\\
\chi_a&=\Lambda_a{}^b(x)\chi_b,
\label{KK'Pconstrchi}
\end{align}
\end{subequations}
for all three types of biKaehler structure.
\pagebreak[1]

For a type $K$ biKaehler structure, 
the action of the biKaehler sigma model (\ref{action}) takes the form
\begin{align}
S_K&=\int_\Sigma\Big\{\frac{1}{2}(ig_{ab}(x)-3K_{ab}(x))
\barpartial x^a\partial x^b-ig^{ab}(x)\overline{\chi}{}_a\chi_b
\label{actionK}
\\
&\hphantom{=\int_\Sigma\Big\{}
+\psi^a\barnabla y_a+\overline{\psi}{}^a\nabla y_a
-R^a{}_{ced}K^{eb}(x)\overline{\psi}{}^c\psi^dy_ay_b\Big\}.\nonumber
\end{align}
The symmetry variations (\ref{deltafields}) of the fields become 
\begin{subequations}
\label{deltafieldsK}
\begin{align}
\delta x^a&=-K^{ab}(x)y_b,\vphantom{\int}
\label{deltaxK}
\\
\delta y_a&=\Gamma^b{}_{ad}K^{dc}(x)y_by_c,\vphantom{\int}
\label{deltayK}
\\
\delta\psi^a&=\Gamma^a{}_{bd}K^{dc}(x)\psi^by_c
+\overline{\Lambda}^a{}_b(x)\partial x^b,\vphantom{\int}
\label{deltapsiK}
\\
\delta\chi_a&=\Gamma^b{}_{ad}K^{dc}(x)\chi_by_c.\vphantom{\int}
\label{deltachiK}
\end{align}
\end{subequations}
For a type $K'$ biKaehler structure, the action (\ref{action}) takes the
simple form
\begin{align}
S_{K'}&=\int_\Sigma i\big(\overline{\chi}{}_a\partial x^a-
\chi_a\barpartial x^a+\overline{\psi}{}^a\nabla y_a
-\psi^a\barnabla y_a\big).
\label{actionK'}
\end{align}
The symmetry variations (\ref{deltafields})  of the fields become
\begin{subequations}
\label{deltafieldsK'}
\begin{align}
\delta x^a&=0,\vphantom{\int}
\label{deltaxK'}
\\
\delta y_a&=0,\vphantom{\int}
\label{deltayK'}
\\
\delta\psi^a&=-i\overline{\Lambda}^a{}_b(x)\partial x^b,\vphantom{\int}
\label{deltapsiK'}
\\
\delta\chi_a&=-i\Lambda_a{}^b(x)\nabla y_b.\vphantom{\int}
\label{deltachiK'}
\end{align}
\end{subequations}
For a type $P$ biKaehler structure, the action (\ref{action}) reads as
\begin{align}
S_P&=\int_\Sigma\Big\{\frac{1}{2}(iQ_{+ab}(x)-3Q_+K_{ab}(x))
\barpartial x^a\partial x^b
-iQ_+{}^{ab}(x)\overline{\chi}{}_a\chi_b
\label{actionP}\\
&\hphantom{=\int_\Sigma\Big\{}
+iQ_-{}^a{}_b(x)(\overline{\chi}{}_a\partial x^b-\chi_a\barpartial x^b)
+Q_+{}^a{}_b(x)
(\psi^b\barnabla y_a+\overline{\psi}{}^b\nabla y_a)\vphantom{\int}\nonumber\\
&\hphantom{=\int_\Sigma\Big\{}
+iQ_-{}^a{}_b(x)
(\overline{\psi}{}^b\nabla y_a-\psi^b\barnabla y_a)
-R^a{}_{ced}Q_+K^{eb}(x)\overline{\psi}{}^c\psi^dy_ay_b\Big\}.\vphantom{\int}
\nonumber
\end{align}
The symmetry variations (\ref{deltafields}) of the fields become 
\begin{subequations}
\label{deltafieldsP}
\begin{align}
\delta x^a&=-Q_+K^{ab}(x)y_b,\vphantom{\int}
\label{deltaxP}
\\
\delta y_a&=\Gamma^b{}_{ad}Q_+K^{dc}(x)y_by_c,\vphantom{\int}
\label{deltayP}
\\
\delta\psi^a&=\Gamma^a{}_{bd}Q_+K^{dc}(x)\psi^by_c
+\overline{\Lambda}(Q_+-iQ_-)^a{}_b(x)\partial x^b,\vphantom{\int}
\label{deltapsiP}
\\
\delta\chi_a&=\Gamma^b{}_{ad}Q_+K^{dc}(x)\chi_by_c
-i\Lambda Q_-{}_a{}^b(x)\nabla y_b.\vphantom{\int}
\label{deltachiP}
\end{align}
\end{subequations}

It is remarkable that, for the $K$, $K'$ and $P$ models, the gauge fermion
$\Psi$ and the topological action $S_{\mathrm{top}}$ entering the basic relation 
(\ref{gaugefermiondef}) (cf. eqs. (\ref{gaugefermion}), (\ref{AB}), 
(\ref{C=J2-P2}), (\ref{Stoptop})) do not depend on the choice of 
the central element $Z^a{}_b$. The gauge fermion is given by 
the same expression for the three models
\begin{equation}
\Psi_K=\Psi_{K'}=\Psi_P=\int_\Sigma\frac{i}{2}
g_{ab}(x)(\overline{\psi}{}^a\partial x^b-\psi^a\barpartial x^b).
\label{gaugefermionKK'P}
\end{equation}
The topological term is given by 
\begin{equation}
S_{\mathrm{top}K}=-\int_\Sigma K_{ab}(x)\barpartial x^a\partial x^b,
\label{StopK}
\end{equation}
for the $K$ model \hphantom{xxxxxxxxxxxxxxxxxx}
\begin{equation}
S_{\mathrm{top}K'}=0,\vphantom{\int}
\label{StopK'}
\end{equation}
for the $K'$ model, and, finally
\begin{equation}
S_{\mathrm{top}P}=-\int_\Sigma Q_+K_{ab}(x)\barpartial x^a\partial x^b,
\label{StopP}
\end{equation}
for the $P$ model. 

Using the fixed point theorem of ref. \cite{Witten3}, or directly from
eqs. (\ref{moduli}), we can write down with little effort
the equations defining the moduli space associated with the $K$, $K'$ and $P$
models. For the $K$ model, they read 
\begin{subequations}
\label{moduliK}
\begin{align}
&\Lambda^a{}_b(x)\barpartial x^b=0,\vphantom{\int}
\label{moduli1K}
\\
&y_b=0.\vphantom{\int}
\label{moduli2K}
\end{align}
\end{subequations}
For the $K$ model, we find
\begin{subequations}
\label{moduliK'}
\begin{align}
&\Lambda^a{}_b(x)\barpartial x^b=0,\vphantom{\int}
\label{moduli1K'}
\\
&\overline{\Lambda}_a{}^b(x)\barnabla y_b=0.\vphantom{\int}
\label{moduli2K'}
\end{align}
\end{subequations}
Finally, for the $P$ model, we have
\begin{subequations}
\label{moduliP}
\begin{align}
&Q_+K^{ab}(x)y_b=0,\vphantom{\int}
\label{moduli1P}
\\
&\Lambda(Q_++iQ_-)^a{}_b(x)\barpartial x^b=0,\vphantom{\int}
\label{moduli2P}
\\
&\overline{\Lambda}Q_-{}_a{}^b(x)\barnabla y_b=0.\vphantom{\int}
\label{moduli3P}
\end{align}
\end{subequations}

Let us discuss the above results. In the $K$ model, 
the field $\chi_a$ is non propagating and decouples from the rest of the
fields in the action, a peculiarity of the $K$ model, which is
not shared by the other biKaehler sigma models. 
Thus, $\chi_a$ may be set to zero \hphantom{xxxxxxxxxxxxxx}
\begin{equation}\label{chi=0}
\chi_a=0.\vphantom{\int}
\end{equation}
This is consistent with the variation (\ref{deltachiK}).
After this is done, by inspection of (\ref{actionK}), (\ref{deltafieldsK}), 
one realizes immediately that $K$ model is nothing but Witten's $A$ 
topological sigma model \cite{Witten1,Witten2} up to a few minor differences. 
In the usual formulation of the $A$ model, instead of the field $y_a$, one uses 
the related field 
\begin{equation}
r^a=-K^{ab}(x)y_b.\vphantom{\int}
\label{ra}
\end{equation}
Its symmetry variation is \hphantom{xxxxxxxxxxxxxx}
\pagebreak[2]
\begin{equation}
\delta r^a=0.\vphantom{\int}
\label{deltara}
\end{equation}
Further, the normalization of the topological term is different from the
one given here. From (\ref{moduliK}), we see that the moduli space 
of the $K$ model classifies holomorphic embeddings $x^a$ of the world sheet 
$\Sigma$ into the target manifold $M$, in agreement with the well 
known property of the $A$ model. 

By inspecting (\ref{actionK'}), it is easy to see that the 
$K'$ model is nothing but the infinite radius limit of 
the $A$ model (in the first order formulism) \cite {Witten1}
(see also \cite{Frenkel}). So, we may call it 
also the $A'$ model. This is a sort of $bc$ system, though, strictly 
speaking, it is not, since the operator $\nabla$ contains a non linear $x^a$
dependence via the connection coefficients $\Gamma^a{}_{bc}$. 
The $A'$ model has another symmetry besides (\ref{deltafieldsK'}).
However, this holds only in the infinite radius limit, in which
non invariance terms proportional to $g^{ab}$ can be neglected. 
(\ref{deltafieldsK'}), conversely, is an exact symmetry. From
(\ref{moduliK'}), it appears the moduli space 
of the $K'$ model classifies pairs $(x^a,y_a)$ constituted by 
a holomorphic embedding of $\Sigma$ into $M$ and a holomorphic 
section of $x^*\Pi T^{*1,0}M$.

The $P$ model interpolates between the $K$ and $K'$ model, to which it
reduces, when $L^a{}_b=\delta^a{}_b,~-\delta^a{}_b$, respectively.
The $P$ model, as far as we know, is not related to any 
known topological sigma model. Also its moduli space is apparently
unknown. 
  
Consider next a biKaehler structure of type $HK$
corresponding to a hyperKaehler structure $g_{ab}$, $K_i{}^a{}_b$
(cf. eqs. (\ref{HK}), (\ref{HKstructure})).
Below, we set 
\begin{equation}\label{Lambda2=Lambda-}
\Lambda_2{}^a{}_b=\Lambda_-{}^a{}_b.
\end{equation}
(cf. eq. (\ref{Pipm})).

For a biKaehler type $HK$ target structure, the fields $\psi^a$, $\chi_a$
satisfy
\begin{subequations}\label{HKconstrpsichi}
\begin{align}
\psi^a&=\overline{\Lambda}_2{}^a{}_b(x)\psi^b,\vphantom{\int}
\label{HKconstrpsi}
\\
\chi_a&=\Lambda_2{}_a{}^b(x)\chi_b.
\label{HKconstrchi}
\end{align}
\end{subequations}

For a type $HK$ biKaehler structure,  the action (\ref{action}) reads as
\begin{align}
S_{HK}&=\int_\Sigma\Big\{\frac{1}{4}(ig_{ab}(x)-5K_{2ab}(x))\barpartial x^a\partial x^b
-ig^{ab}(x)\overline{\chi}{}_a\chi_b
\label{HKaction}
\\
&\hphantom{=\int_\Sigma\Big\{}
+S^{ab}(x)
\big(\overline{\chi}{}_a+g_{ac}(x)\barpartial x^c\big)
\big(\chi_b+g_{bd}(x)\partial x^d\big)\vphantom{\int}
\nonumber
\\
&\hphantom{=\int_\Sigma\Big\{}
+\frac{1}{2}(1+2S)^a{}_b(x)\big(\psi^b\barnabla y_a+\overline{\psi}{}^b\nabla y_a\big)
+R^a{}_{ced}P^{eb}(x)\overline{\psi}{}^c\psi^dy_ay_b\Big\},\vphantom{\int}
\nonumber
\end{align}
where the tensors $P^a{}_b$, $S^a{}_b$ are given by  
\begin{subequations}
\label{HKPS}
\begin{align}
P^a{}_b&=\frac{1}{2}(K_1-K_2)^a{}_b, 
\label{HKP}
\\
S^a{}_b&=\frac{1}{2}(K_1+K_2+K_3)^a{}_b
\label{HKS}
\end{align}
\end{subequations}
The symmetry variations (\ref{deltafields})  of the fields become
\begin{subequations}
\label{HKdeltafields}
\begin{align}
\delta x^a&=P^{ab}(x)y_b,\vphantom{\int}
\label{HKdeltax}
\\
\delta y_a&=-\Gamma^b{}_{ad}P^{dc}(x)y_by_c,\vphantom{\int}
\label{HKdeltay}
\\
\delta\psi^a&=-\Gamma^a{}_{bd}P^{dc}(x)\psi^by_c
+\Big(\frac{1}{2}\overline{\Lambda}{}_2{}^a{}_b(x)+F_-{}^a{}_b(x)\Big)\partial x^b
+G_-{}^{ab}(x)\chi_b,\vphantom{\int}
\label{HKdeltapsi}
\\
\delta\chi_a&=-\Gamma^b{}_{ad}P^{dc}(x)\chi_by_c
+F_+{}^b{}_a(x)\nabla y_b
-R^b{}_d{}^c{}_eG_+{}^e{}_a(x)\psi^dy_by_c.\vphantom{\int}
\label{HKdeltachi}
\end{align}
\end{subequations}
where the tensors $F_\pm{}^a{}_b$, $G_\pm{}^a{}_b$ are given by 
\begin{subequations}
\label{HKAB}
\begin{align}
F_{\pm}{}^a{}_b&=\frac{1}{4}(K_1+K_2-i\pm iK_3)^a{}_b,
\label{HKA}
\\
G_{\pm}{}^a{}_b&=\frac{1}{4}(K_1-K_2+i\pm iK_3)^a{}_b.
\label{HKB}
\end{align}
\end{subequations}

For the $HK$ sigma model, the gauge fermion $\Psi$ and the topological action 
$S_{\mathrm{top}}$ entering relation (\ref{gaugefermiondef}) 
(cf. eqs. (\ref{gaugefermion}), (\ref{AB}), 
(\ref{C=J2-P2}), (\ref{Stoptop})) are given by
\begin{align}
\Psi_{HK}&=\int_\Sigma\frac{i}{2}\Big[\big(g_{ab}(x)+H_1K_{3ab}(x)\big)
(\overline{\psi}{}^a\partial x^b-\psi^a\barpartial x^b)~~~~~~~~~~
\label{gaugefermionHK}\\
&\hskip 5.5cm
+H_2{}^a{}_b(x)(\overline{\chi}_a\psi^b-\chi_a\overline{\psi}{}^b)\Big],\vphantom{\int}
\nonumber
\end{align}
where the tensors $H_1{}^a{}_b$, $H_2{}^a{}_b$ are given by 
\begin{equation}
H_1{}^a{}_b=\frac{1}{18}(K_1-17K_2-4K_3)^a{}_b,\qquad H_2{}^a{}_b=\frac{1}{18}(K_1+K_2+5K_3)^a{}_b,
\label{Cpm}
\end{equation}
and \hphantom{xxxxxxxxxxxxxx}
\begin{equation}
S_{\mathrm{top}HK}=\int_\Sigma 2P_{ab}(x)\barpartial x^a\partial x^b.
\label{StopHK}
\end{equation}
The expression of $\Psi_{HK}$ is not particularly illuminating, though its
complexity may be a cohomological artifact.

The moduli space of the $HK$ model is defined by the equations
\begin{subequations}\label{moduliHK}
\begin{align}
&\Big(\frac{1}{2}\Lambda_2{}^a{}_b(x)+\overline{F}{}_-{}^a{}_b(x)\Big)\barpartial x^b
+\overline{G}{}_-{}^{ab}(x)\overline{\chi}{}_b=0,\vphantom{\int}\label{moduliHK1}\\
&~~y_a=0\vphantom{\int}\label{moduliHK2}
\end{align}
\end{subequations}
obtainable for instance from (\ref{moduli}), (\ref{y=0}). 
These equations are characterized by the explicit appearance of the field
$\chi_a$, in contrast to what happens for the $K$, $K'$ and $P$ models. 

To the best of our knowledge, the $HK$ model is not related to any 
known topological sigma model. Also the associated moduli space
is apparently unknown. The fact that the field $\chi_a$ appears explicitly in 
the moduli space equations (\ref{moduliHK}) may indicate that it plays a role 
rather different from that it does in the $K$, $K'$ and $P$ models.
At this stage, it is difficult to assess the relevance and even the
consistency of the $HK$ model.

\vfill\eject

\begin{small}
\section{\bf Relation to the Hitchin model}
\label{sec:Hitchin}
\end{small}
In this section, we review briefly the Hitchin sigma model, worked out in
refs. \cite{Zucchini1,Zucchini2}, restricting ourselves to the case 
where the target space generalized
complex structure is actually a generalized Kaehler structure
\cite{Hitchin,Gualtieri}. We then show
that the action and the symmetries of the biKaehler sigma model can be obtained 
by gauge fixing the Batalin--Vilkovisky master action of the Hitchin 
model by restricting to a suitably chosen submanifold of field space, that
is Lagrangian with respect to the Batalin--Vilkovisky odd symplectic form. 

In general, the fields of a 2--dimensional field theory are differential 
forms on a oriented $2$--dimensional manifold $\Sigma$. They can be viewed 
as elements of the space $\mathrm{Fun}(\Pi T\Sigma)$ of functions on the parity 
reversed tangent bundle $\Pi T\Sigma$ of $\Sigma$, which we shall call 
de Rham superfields. More explicitly, we associate with the coordinates $t^\alpha$ of 
$\Sigma$ Grassmann odd partners $\tau^\alpha$ with $\deg t^\alpha=0$, $\deg\tau^\alpha=1$.
A de Rham superfield $\ul \psi(t,\tau)$ is a triplet 
formed by a $0$--, $1$--, $2$--form field $\psi^{(0)}(t)$, 
$\psi^{(1)}{}_\alpha(t)$, $\psi^{(2)}{}_{\alpha\beta}(t)$ organized as
\begin{equation}
\ul \psi(t,\tau)=\psi^{(0)}(t)+\tau^\alpha\psi^{(1)}{}_\alpha(t)
+\frac{1}{2}\tau^\alpha\tau^\beta\psi^{(2)}{}_{\alpha\beta}(t).
\end{equation}
The forms $\psi^{(0)}$, $\psi^{(1)}$, $\psi^{(2)}$ are called the 
de Rham components of $\ul\psi$. 

$\Pi T\Sigma$ is endowed with a natural differential $\ul d$ defined by 
\begin{equation}
\ul dt^\alpha=\tau^\alpha,\qquad \ul d\tau^\alpha=0.\vphantom{\Big[}
\end{equation}
In this way, the exterior differential $d$ of $\Sigma$ 
can be identified with the operator 
\begin{equation}
\ul d=\tau^\alpha\partial_\alpha.
\end{equation}

The coordinate invariant integration measure of $\Pi T\Sigma$ is 
\begin{equation}
\mu={\rm d}t^1{\rm d}t^2{\rm d}\tau^1{\rm d}\tau^2.
\end{equation}
Any de Rham superfield $\ul \psi$ can be integrated on $\Pi T\Sigma$ according to
the prescription
\begin{equation}
\int_{\Pi T\Sigma}\mu\,\ul\psi=\int_\Sigma\frac{1}{2}
{\rm d}t^\alpha {\rm d}t^\beta\psi^{(2)}{}_{\alpha\beta}(t).
\end{equation}

The components of the relevant de Rham superfields carry, besides the form
degree, also a ghost degree. We shall limit ourselves 
to homogeneous superfields, that is superfields $\ul \psi$ for which 
the sum of the form and ghost degree is the same for the three components 
$\psi^{(0)}$, $\psi^{(1)}$, $\psi^{(2)}$ of $\ul\psi$. 
The common value of that sum is the superfield (total) degree $\deg\ul \psi$. 
It is easy to see that the differential operator $\ul d$ and the
integration operator $\int_{\Pi T\Sigma}\mu$ carry 
degree $1$ and $-2$, respectively. 

It is often necessary to choose a complex structure on
$\Sigma$. With this, there are associated complex coordinates
for $\Sigma$, $z$, and their Grassmann odd partners,
$\zeta$, and their complex conjugates. As before, $\deg z=0$, $\deg\zeta=1$. 
All the above relations can be written in terms these 
coordinates, if one wishes so. Further, once a complex structure is given,
we can define the Cauchy--Riemann operator
\begin{equation}\label{holpartial}
\ul \partial=\zeta\partial_z
\end{equation}
and its complex conjugate and, with this, a notion of
holomorphy for superfields. $\ul \partial$ has obviously degree $1$. 

Now, we shall introduce the Hitchin sigma model \cite{Zucchini1,Zucchini2}.
The basic fields of the model are a degree $0$
superembedding $\ul x\in\Gamma(\Pi T\Sigma, M)$ and a degree $1$
supersection $\ul y\in\Gamma(\Pi T\Sigma, x^*\Pi T^*M)$, where $\Pi T^*M$
is the parity reversed cotangent bundle of $M$.
With respect to each local coordinate 
of $M$, $\ul x$, $\ul y$ are given as de Rham superfields $\ul x^a$, $\ul y_a$. 
The Batalin--Vilkovisky odd symplectic form is 
\begin{equation}\label{BVOmega}
\Omega_{BV}=\int_{\Pi T\Sigma}\mu\,\delta \ul x^a\delta \ul y_a,
\end{equation}
where, here, $\delta$ denotes the differential operator in field space.  
$\Omega_{BV}$ is a closed functional form, $\delta\Omega_{BV}=0$. In this way, 
one can define Batalin--Vilkovisky 
antibrackets $(\,,)_{BV}$ in standard fashion by the 
formula: 
\begin{equation}\label{BVbrackets}
(F,G)_{BV}=\int_{\Pi T\Sigma}\mu\,\bigg[
\frac{\delta_r F}{\delta \ul x^a}\frac{\delta_l G}{\delta \ul y_a}
-\frac{\delta_r F}{\delta \ul y_a}\frac{\delta_l G}{\delta \ul x^a}\bigg],
\end{equation}
for any two functionals $F$, $G$ of $\ul x^a$, $\ul y_a$, where the subscripts
$l$, $r$ denote left, right functional differentiation, respectively.

In the Hitchin sigma model, the target space geometry is specified by a 
generalized complex structure $\mathcal{J}^A{}_B$ \cite{Hitchin}. 
In the case under our study, the structure $\mathcal{J}^A{}_B$ is the
generalized Kaehler structure corresponding to a biKaehler structure $g_{ab}$,
$K_\pm{}^a{}_b$ \cite{Gualtieri} and, in block form, is given by
\begin{equation}\label{calJ}
{\cal J}^A{}_B=\bigg(\begin{matrix}J^a{}_b& P^{ab}\\ P_{ab}& J_a{}^b\end{matrix}\bigg),
\end{equation}
where the tensors $J^a{}_b$, $P^a{}_b$ are given by (\ref{JP}). 
The action of the associated Hitchin model is 
\begin{equation}\label{BVaction}
S_{GK}=\int_{\Pi T\Sigma}\mu\,\Big[\frac{1}{2}P^{ab}(\ul x)\ul y_a\ul y_b
+J^a{}_b(\ul x)\ul y_a\ul d\ul x^b+\frac{1}{2}P_{ab}(\ul x)\ul d\ul x^a\ul d\ul x^b\Big].
\end{equation}
Actually, in the Hitchin sigma model, as formulated in \cite{Zucchini1, Zucchini2},
the action $S_{GK}$ contains an extra term $\int_{\Pi T\Sigma}\mu\,\ul y_a\ul d\ul x^a$
absent here. It is straightforward to see that this omission does not
alter the main property of the model, that is the correspondence between the
integrability conditions of the generalized almost complex structure $\cal J^A{}_B$ 
and the restrictions on target space geometry implied by the Batalin--Vilkovisky 
classical master equation. Indeed, it can be checked that 
$S_{GK}$ satisfies the classical Batalin--Vilkovisky master equation 
\begin{equation}\label{masterBV}
(S_{GK},S_{GK})_{BV}=0,
\end{equation}
as a consequence of (\ref{K2=-1}), (\ref{K+Kt=0}), (\ref{nablaK=0}).

The Batalin--Vilkovisky variations $\delta_{BV}\ul x^a$, 
$\delta_{BV}\ul y_a$ are defined by 
\begin{subequations}\label{deltaBVxydef}
\begin{align}
\vphantom{\int} 
\delta_{BV}\ul x^a&=(S_{GK},\ul x^a)_{BV},\label{deltaBVxdef}\\
\vphantom{\int}
\delta_{BV}\ul y_a&=(S_{GK},\ul y_a)_{BV}.\label{deltaBVydef}
\end{align}
\end{subequations}
Using (\ref{BVaction}), (\ref{deltaBVxydef}), it is straightforward to obtain the 
explicit expressions of $\delta_{BV} \ul x^a$, $\delta_{BV} \ul y_a$
\begin{subequations}\label{deltaBVxy}
\begin{align}
\vphantom{\int}
\delta_{BV} \ul x^a&=P^{ab}(\ul x)\ul y_b+J^a{}_b(\ul x)\ul d\ul x^b,\label{deltaBVx}\\
\vphantom{\int}
\delta_{BV} \ul y_a&=-\Gamma^b{}_{ad}P^{dc}(\ul x)\ul y_b\ul y_c
-\Gamma^b{}_{ad}J^d{}_c(\ul x)\ul y_b\ul d\ul x^c
+J^b{}_a(\ul x)\ul \nabla\ul y_b, \label{deltaBVy}
\end{align}
\end{subequations}
where \hphantom{xxxxxxxxxxxxxxxxx}
\begin{equation}\label{realnabla}
\ul \nabla =\ul d\pm\Gamma^{\mathbf{\cdot}}{}_{\mathbf{\cdot}a}(\ul x)\ul d\ul x^a.
\end{equation}
The operator $\delta_{BV}$ has degree $+1$.
As is well known, the master equation (\ref{masterBV}) implies that
$\delta_{BV}$ is nilpotent \hphantom{xxxxxxxxxxxxxxxxx}
\begin{equation}\label{deltaBV2=0}
\delta_{BV}{}^2=0.
\end{equation}
The associated cohomology is the classical Batalin--Vilkovisky cohomology.
Also, by (\ref{masterBV}), one has \hphantom{xxxxxxxxxxxxxxxxx}
\begin{equation}\label{deltaBVSGK=0}
\delta_{BV} S_{GK}=0.
\end{equation}

As it is, the action $S_{GK}$ is not suitable for quantization because it possesses
a gauge symmetry as a consequence of (\ref{deltaBVSGK=0}).
This gauge symmetry renders the kinetic terms of the
fields ill defined. Gauge fixing is required. 
(We refer the reader to ref. \cite{Gomis} for an exhaustive treatment of gauge fixing 
in the framework of the Batalin--Vilkovisky quantization algorithm.) 
As is well known, this is
carried out by restricting the action to a field space submanifold 
$\mathfrak{L}$, that is Lagrangian with respect to  
the Batalin--Vilkovisky odd symplectic form $\Omega_{BV}$.
The resulting quantum field theory does not depend on the choice of 
$\mathfrak{L}$ for continuous deformation of the latter.
However, not every choice of $\mathfrak{L}$ leads to a well defined quantum field theory.
A particular choice of $\mathfrak{L}$, then, can be justified only {\it a posteriori}.
Below, we shall implement the gauge fixing following closely the methodology of 
Alexandrov, Kontsevich, Schwartz and Zaboronsky \cite{AKSZ}, with which they
worked out a formulation of the $A$ topological sigma model {\it \`a la} Batalin--Vilkovisky.

The definition of $\mathfrak{L}$ requires the choice of a complex structure on
$\Sigma$. With this given, we define the differential operator
\begin{equation}\label{holD}
\ul D=\zeta\partial_\zeta\pm\Gamma^{\mathbf{\cdot}}{}_{\mathbf{\cdot}a}(\ul x)
\zeta\partial_\zeta\ul x^a
\end{equation}
and its complex conjugate, of degree $0$. $\ul D$ turns out to be very useful
because of its remarkable properties. $\ul D$ is a projector, as
\begin{equation}\label{D2=D}
\ul D^2=\ul D,\vphantom{\int}
\end{equation}
as is easy to check. Further, one has
\begin{equation}\label{intDbarD}
\int_{\Pi T\Sigma}\mu\,\ul \psi=\int_{\Pi T\Sigma}\mu\,\ul{\overline{D}}\ul D\ul \psi,
\end{equation}
for any superfield $\ul \psi$ of the our sigma model. 

The field space submanifold $\mathfrak{L}$ is defined by the constraints
\begin{subequations}\label{lagrangexy}
\begin{align}
&\Lambda_-{}^a{}_b(\ul x)\ul D\ul x^b\simeq 0,\vphantom{\int} \label{lagrangex}\\
&\overline{\Lambda}_-{}_a{}^b(\ul x)\ul D\big[\ul y_b+ig_{bc}(\ul x)(\ul \partial \ul x^c
+\ul \barpartial \ul x^c)\big]\simeq 0,\vphantom{\int}\label{lagrangey} 
\end{align}
\end{subequations}
where $\Lambda_-{}^a{}_b$ is the projector (\ref{Pipm}). 
Here and below, the symbol $\simeq$ denotes equality holding upon restriction
to $\mathfrak{L}$ in field space. By direct verification, one can show 
that the Batalin--Vilkovisky odd symplectic form vanishes on $\mathfrak{L}$, 
\begin{equation}\label{BVOmega=0}
\Omega_{BV}\simeq 0. 
\end{equation}
In more precise terms, it is the the pull--back of $\Omega_{BV}$ by the injection
$\iota_{\mathfrak{L}}$ of $\mathfrak{L}$ into field space, $\iota_{\mathfrak{L}}{}^*\Omega_{BV}$, 
that vanishes. Thus, $\mathfrak{L}$ is a field space Lagrangian submanifold for the 
Batalin--Vilkovisky symplectic form $\Omega_{BV}$, as desired.
The verification of (\ref{BVOmega=0}) is straightforward though lengthy.
Here, we shall provide a few usefull hints about the way it is carried out.
To controll covariance, one rewrites (\ref{BVOmega}) conveniently as
\hphantom{xxxxxxxxxxxxxxxxxxxxxxxxx}
\begin{equation}\label{}
\Omega_{BV}=\int_{\Pi T\Sigma}\mu\,\delta \ul x^a\delta_{\mathrm{c}} \ul y_a,  
\end{equation}
where the covariant variations $\delta_{\mathrm{c}} \ul y_a$ is given by 
\begin{equation}\label{}
\delta_{\mathrm{c}} \ul y_a=\delta \ul y_a-\Gamma^b{}_{ac}(\ul x)\delta\ul x^c \ul y_b.
\end{equation}
Using (\ref{intDbarD}), one has then
\begin{equation}\label{}
\Omega_{BV}=\int_{\Pi T\Sigma}\mu\,\big[
\ul{\overline{D}}\ul D\delta \ul x^a\delta_{\mathrm{c}} \ul y_a  
+\ul{\overline{D}}\delta \ul x^a\ul D\delta_{\mathrm{c}} \ul y_a  
+\ul D\delta \ul x^a\ul{\overline{D}}\delta_{\mathrm{c}} \ul y_a  
+\delta \ul x^a\ul{\overline{D}}\ul D\delta_{\mathrm{c}} \ul y_a  
\big].
\end{equation}
By applying the operator $\delta$ to the constraints (\ref{lagrangexy}), 
one obtains relations involving $\ul D\delta \ul x^a$, $\ul D\delta_{\mathrm{c}} \ul y_a$,
$\ul{\overline{D}}\ul D\delta \ul x^a$, $\ul{\overline{D}}\ul D\delta_{\mathrm{c}} \ul y_a$
and their complex conjugates, 
which, together with (\ref{lagrangexy}),  allow one to show (\ref{BVOmega=0}).

Using (\ref{intDbarD}), it is simple to show that 
\begin{align}
S_{GK}&=\int_{\Pi T\Sigma}\mu\,\Big[P^{ab}(\ul x)\big(\ul {\overline{D}}\ul y_a\ul D\ul y_b
+\ul y_a\ul{\overline{D}}\ul D\ul y_b\big)+P_{ab}(\ul x)\ul\barpartial\ul x^a\ul\partial\ul x^b
\label{SGKexplicit}\\
&\hphantom{\int_{\Pi T\Sigma}\mu\,\Big[}
\,+J^a{}_b(\ul x)\big(\ul D\ul y_a\ul\barpartial\ul x^b
+\ul{\overline{D}}\ul y_a\ul\partial\ul x^b
+y_a(\ul{\overline{\nabla}}\ul D\ul x^b+\ul\nabla\ul{\overline{D}}\ul x^b\big)\Big],
\nonumber
\end{align}
where $\ul\nabla$ is the covariant Cauchy--Riemann operator 
\begin{equation}\label{holnabla}
\ul \nabla=\ul \partial\pm\Gamma^{\mathbf{\cdot}}{}_{\mathbf{\cdot}a}(\ul x)\ul\partial\ul x^a
\end{equation}
and similarly for its complex conjugate \footnote{\vphantom{$\bigg[$} Here, we are
changing our notation with respect to (\ref{realnabla}).}. 
Let us call a superfield $\ul\phi$ of one of the forms $\ul\phi=\ul D\ul\psi$,
$\ul\phi=\ul{\overline{D}}\ul\psi$, $\ul\phi=\ul{\overline{D}}\ul D\ul\psi$ 
a descendent of $\ul\psi$. 
(\ref{SGKexplicit}) shows that $S_{GK}$ is a functional of the 
superfields $\ul x^a$, $\ul y_a$ both explicitly and implicitly 
through their descendent superfields $\ul D\ul x^a$,
$\ul D\ul y_a$ and $\ul{\overline{D}}\ul D\ul y_a$ and their complex conjugates.
On the Lagrangian submanifold $\mathfrak{L}$, the superfields $\ul D\ul x^a$,
$\ul D\ul y_a$ and $\ul{\overline{D}}\ul D\ul y_a$ satisfy certain 
relations entailed by (\ref{lagrangexy}). 
A detailed analysis shows that these relations allow one to express   
$\ul D\ul x^a$, $\ul D\ul y_a$ and $\ul{\overline{D}}\ul D\ul y_a$ 
in terms of the superfields $\ul x^a$, $\ul y_a$ and the further superfields 
\begin{subequations}\label{ulpsichi}
\begin{align}
\ul\psi^a&=\overline{\Lambda}{}_-{}^a{}_b(\ul x)\ul D\ul x^b,\vphantom{\int}\label{ulpsi}\\
\ul\chi_a&=\Lambda_-{}_a{}^b(\ul x)\ul D\ul y_b,\vphantom{\int}\label{ulchi} 
\end{align}
\end{subequations}
explicitly, i. e. without any appearance of their descendent superfields, such as 
$\ul D\psi^a$, $\ul D\ul\chi_a$, etc.:
\begin{subequations}\label{ulDbarDxy}
\begin{align}
\ul D\ul x^a&\simeq\ul\psi^a,\vphantom{\int}\label{ulDx}\\
\ul D\ul y_a&\simeq\ul\chi_a
-i\overline{\Lambda}{}_{-ab}(\ul x)(\ul\partial \ul x^b+\ul{\overline{\nabla}}\ul\psi^b),
\vphantom{\int}\label{ulDy}\\
\ul{\overline{D}}\ul D\ul y_a&\simeq
-R^b{}_{ecd}\overline{\Lambda}{}_-{}^e{}_a(\ul
x)\ul{\overline{\psi}}^c\ul\psi^d\ul y_b
+i(\Lambda_--\overline{\Lambda}{}_-)_{ab}(\ul x)
(\ul{\overline{\nabla}}\ul\psi^b+\ul\nabla\ul{\overline{\psi}}^b).
\vphantom{\int}\label{DbarDy}
\end{align}
\end{subequations}
By (\ref{SGKexplicit}), we have therefore, \hphantom{xxxxxxxxxxxxxxxxxxxxxxxxx}
\begin{equation}\label{Sgf}
S_{GK}\simeq S_{GK}^{\mathrm{gf}},
\end{equation}
where $S_{GK}^{\mathrm{gf}}$ is an explicit functional of the superfields 
$\ul x^a$, $\ul y_a$, $\ul \psi^a$, $\ul\chi_a$. From (\ref{SGKexplicit}),
it appears that $S_{GK}^{\mathrm{gf}}$ depends only on the lowest non zero de 
Rham components of the superfields $\ul x^a$, $\ul y_a$, $\ul \psi^a$,
$\ul\chi_a$, which we denote $x^a$, $y_a$, $\psi^a$, $\chi_a$.
These are precisely the fields of the biKaehler sigma model
introduced in sect. \ref{sec:sigma}.
From (\ref{ulpsichi}), it is evident that  $\psi^a$, $\chi_a$ obey the
constraints (\ref{constrpsichi}). Through a detailed calculation, one finds 
further that, in terms of $x^a$, $y_a$, $\psi^a$, $\chi_a$, 
$S_{GK}^{\mathrm{gf}}$ equals the biKaehler sigma model action $S$ given in eq.
(\ref{action}). Similarly, one can derive from the Batalin--Vilkovisky variations 
(\ref{deltaBVxy}) the symmetry variations (\ref{deltafields}). 
In this way, we were able to show the relation of the biKaehler sigma model to
the Hitchin sigma model for generalized Kaehler target. 

The result we have just obtained is interesting in itself, but it is also
interesting because of the light it sheds on the nature of world sheet 
and target space geometrical data, on which the quantum field theory 
associated with biKaehler sigma effectively depends. We obtained 
the biKaehler sigma model by gauge fixing the Hitchin sigma model 
with generalized Kaehler target following the general prescriptions
of Batalin--Vilkovisky formalism. We know in this way that the resulting 
gauge fixed field theory depends generically on the geometrical data contained 
in the Hitchin sigma model action $S_{GK}$, but it is independent from those
defining the Lagrangian submanifold $\mathfrak{L}$ \cite{AKSZ}.
Now, the action $S_{GK}$ has the following structure \hphantom{xxxxxxxxxxxxxxxxxxxxxxxxx}
\begin{equation}\label{BVactionsplit}
S_{GK}=S_{GK1}+S_{GK2},
\end{equation}
where $S_{GK1}$, $S_{GK2}$ are given by 
\begin{subequations}\label{SGK12}
\begin{align}
S_{GK1}&=\int_{\Pi T\Sigma}\mu\,\Big[\frac{1}{2}P^{ab}(\ul x)\ul y_a\ul y_b
+J^a{}_b(\ul x)\ul y_a\ul d\ul x^b\Big],
\label{SGK1}\\
S_{GK2}&=\int_{\Pi T\Sigma}\mu\,\frac{1}{2}P_{ab}(\ul x)\ul d\ul x^a\ul d\ul x^b. 
\label{SGK2}
\end{align}
\end{subequations}
Since $P_{ab}$ is a closed $2$--form, $S_{GK2}$ is just a topological
term. As we remarked in sect. \ref{sec:sigma}, terms of this type do not affect 
the field equations and are invariant under all infinitesimal symmetries.
Their values characterize the topological sectors of the field theory,
but such terms in themselves do not affect in any way the quantum structure 
of the field theory. Thus, they may be adjusted as one wishes
as a matter of definition of the model without really changing its quantum 
properties in any essential way. The truly quantum sector of the field 
theory stems, upon gauge fixing, from $S_{GK1}$. This depends only on the
combinations $J^a{}_b$, $P^{ab}$ (cf. eq. (\ref{JP})) of the target space 
geometrical data $g_{ab}$, $K_\pm{}^a{}_b$. The Lagrangian submanifold 
$\mathfrak{L}$ depends also on the chosen complex structure of $\Sigma$
and on $g_{ab}$, $K_-{}^a{}_b$, but, for the reasons recalled above, 
the gauge fixed field theory will not. So, we conclude that
the quantum field theory associated to the biKaehler model
depends effectively only on the combinations $J^a{}_b$, $P^{ab}$ 
of target space biKaehler geometrical data. This solves the problem posed in sect. 
\ref{sec:exactenss}. For the $A$ model, $J^a{}_b=0$, $P^{ab}=-K^{ab}$. We
recover in this way the well known result that the $A$ model depends only on
the target space Kaehler structure.

In his thesis \cite{Gualtieri}, Gualtieri showed that a biKaehler geometry $g_{ab}$,
$K_\pm{}^a{}_b$ is fully equivalent to a pair of commuting generalized complex
structures
\begin{equation}\label{calJ12}
{\cal J}_1{}^A{}_B=\bigg(\begin{matrix}J^a{}_b& P^{ab}\\ P_{ab}& J_a{}^b\end{matrix}\bigg),
\qquad
{\cal J}_2{}^A{}_B=\bigg(\begin{matrix}P^a{}_b& J^{ab}\\ J_{ab}& P_a{}^b\end{matrix}\bigg),
\end{equation}
whose product $-{\cal J}_1{}^A{}_C{\cal J}_2{}^C{}_B$ satisfies a certain
positivity condition. 
The structure ${\cal J}_1{}^A{}_B$ equals the structure ${\cal J}^A{}_B$ of
eq. (\ref{calJ}) in terms of which the Hitchin model action $S_{GK}$ is
defined. So, we could reformulate the above results saying that the quantum
field theory of the biKaehler model depends only on ${\cal J}_1{}^A{}_B$, 
a fact implicit in the results previously obtained 
by several authors \cite{Zabzine1,Kapustin1,Kapustin2,Chiantese}.

\vfill\eject

\begin{small}
\section{\bf Discussion}
\label{sec:discussion}
\end{small}

In this final section, we discuss our results by comparing them with those
of other studies, which have appeared in the literature, and by listing 
the open problems. 

{\it Twisting of the (2,2) supersymmetric biKaehler sigma model}

We recall \cite{Gates} that, for any target space $M$ with biKaehler structure $g_{ab}$,
$K_\pm{}^a{}_b$, there exists a $(2,2)$ supersymmetric sigma model. 
Its action $S$ may contain a closed $2$--form $b_{ab}$, defining a topological
term. Explicitly, in $(1,1)$ superspace notation, $S$ is given by
\begin{equation}
\label{susyaction}
S \;=\; \frac{1}{2} \int d^2\sigma d^2\theta 
\big(g_{ab}(X)+b_{ab}(X)\big)D_+X^a D_-X^b, 
\end{equation}
where $X^a$ is the $(1,1)$ superfield
\begin{equation}
\label{susyfield}
X^a =x^a+\theta^+\psi_+^a+\theta^-\psi_-^a+\theta^-\theta^+F^a,
\end{equation}
defining a superembedding of the the $(1,1)$ superworldsheet into $M$, 
and the $D_\pm$ are the supercovariant derivatives
\begin{equation}
\label{susyDpm}
D_\pm= \frac{\partial}{\partial\theta^\pm}+i\theta^\pm\partial_\pm,  
\qquad \partial_\pm\equiv\partial_0\pm\partial_1.
\end{equation}
The complex structures $K_\pm{}^a{}_b$ do not appear in the action, but they
enter into the definition of the $(2,2)$ supersymmetry variations:
\begin{equation}
\label{susyvars}
\delta X^a=\epsilon^+Q_+X^a +\epsilon^-Q_-X^a 
+\tilde\epsilon{}^+K_+(X)^a{}_b\tilde{Q}_+X^b + \tilde\epsilon{}^- K_-(X)^a{}_b
\tilde{Q}_-X^b,
\end{equation}
where the $\epsilon^\pm$, $\tilde\epsilon{}^\pm$ are anticommuting parameters
and 
\begin{equation}
\label{susyQpm}
Q_\pm= \frac{\partial}{\partial\theta^\pm}-i\theta^\pm\partial_\pm, \qquad
\tilde{Q}_\pm= \frac{\partial}{\partial\theta^\pm}+i\theta^\pm\partial_\pm.
\end{equation}

In refs. \cite{Kapustin1,Kapustin2}, Kapustin and Li proposed a twisting 
prescription to generate a generalized $A$ topological sigma model from the 
$(2,2)$ sigma model. After the topological twist, the fields 
\begin{equation}
r_+{}^a=\overline{\Lambda}{}_+{}^a{}_b(x)\psi_+{}^b,\qquad 
r_+{}^a=\Lambda_-{}^a{}_b(x)\psi_-{}^b
\end{equation}
become $0$--form sections of $x^*\Pi T_+^{0,1}M$, $x^*\Pi T_-^{1,0}M$,
while the fields
\begin{equation}
\phi_+{}^a=\Lambda_+{}^a{}_b(x)\psi_+{}^b,\qquad 
\phi_-{}^a=\overline{\Lambda}{}_-{}^a{}_b(x)\psi_-{}^b 
\end{equation}
become $(0,1)$--, $(1,0)$--form sections of $x^*\Pi T_+^{1,0}M$, $x^*\Pi T_-^{0,1}M$,  
respectively \footnote{$\vphantom{\bigg[}$ The analysis of
 ref. \cite{Kapustin2} is actually broader in scope, as it covers the more 
general case of a biHermitian target space and also considers a generalized $B$ model.}. 
This prescription is rather natural, since, as (\ref{susyvars}) shows, the complex
structures $K_\pm{}^a{}_b$ correspond to the two world sheet chiralities. 
It is clear that this field content cannot match the one of the biKaehler
model studied in this paper (cf. sect. \ref{sec:sigma}). 
Further, according to the same authors, the topological variations of the fields  
$x^a$, $r_\pm{}^a$ are given by
\begin{subequations}
\begin{align}
\delta x^a&=r_+{}^a+r_-{}^a,\\
\delta r_\pm{}^a&=-\Gamma^a{}_{bc}(x)r_\mp{}^br_\pm{}^c,
\end{align}
\end{subequations}
with
\begin{equation}
\delta^2=0.
\end{equation}
The local observables of the theory are thus of the form
\begin{equation}
\widehat{f}=\sum_{p,q\geq 0}\frac{1}{p!q!}
f_{a_1\cdots a_p;b_1\cdots b_q}(x)r_+{}^{a_1}\cdots
r_+{}^{a_p}r_-{}^{b_1}\cdots r_-{}^{b_q},
\end{equation}
where the $f_{a_1\cdots a_p;b_1\cdots b_q}$ belong to $\Omega_+^{0,p}(M)\otimes\Omega_-^{q,0}(M)$
and satisfy
\begin{equation}
p\overline{\Lambda}_+{}^c{}_{[a_1}\nabla_{|c|}f_{a_2\cdots a_p];b_1\cdots b_q}
+(-1)^pq\Lambda_-{}^c{}_{[b_1} \nabla_{|c}f_{a_1\cdots a_p|;b_2\cdots b_q]}=0.
\end{equation}
The associated cohomology has no apparent relation with the
Poisson--Licnerowicz cohomology found in sect. \ref{sec:cohomology}
\footnote{$\vphantom{\bigg[}$ In ref. \cite{Kapustin2}, it is shown that this cohomology is the
total cohomology of a double cochain complex}.

So, seemingly, the construction of Kapustin and Li and ours are fundamentally 
different. As a consequence, our biKaehler sigma model has no immediate interpretation
as a twisted form of a $(2,2)$ supersymmetric sigma model. Of course, one
could envisage other twisting prescriptions, though the one formulated by
Kapustin and Li seems to be the most natural. But this would hardly solve the
discrepancy: indeed, there is no field in the $(2,2)$ supersymmetric sigma
model, which, under twisting, may turn into the $\chi_a$ field of the
biKaehler sigma model. As we have seen, $\chi_a$ decouples in the $A$
model, but it does not so in general. 
 
{\it BiKaehler moduli space}

The biKaehler sigma model is a topological field theory of cohomological type.
It is known that any field theory like that describes the intersection
theory of some moduli space in terms of local quantum field theory. 
On general grounds, the moduli space can be defined as the space of 
solutions of a set of equations, obtainable using the fixed point theorem of 
ref. \cite{Witten3}. For the biKaehler sigma model, we derived these
equations in sect. \ref{sec:exactenss}, see eqs. (\ref{moduli}). 
However, we still do not have any geometrical interpretation 
or analytic understanding of this moduli space 
in general. Moduli spaces are notoriously subtle geometrical--topological
structures. In order to be able to define intersection theory, one needs 
to compactify them and there is no unique of doing that in general. 
Moreover, they are usually plagued by singularities, which render them hardly
amenable by standard means of analysis. 
A detailed investigation of these matters would be required. 

{\it BiKaehler topological sectors and quantum cohomology}

As is well known, the observables of $A$ topological sigma model 
form a ring that is isomorphic to a deformation of the de Rham
cohomology ring, going under the name of quantum cohomology
\cite{Vafa,Witten4,Aspinwall}.
This turns out to be an important invariant in symplectic geometry.
The deformation is made possible by the fact that the model
possesses non trivial topological sectors, with which there are 
associated world sheet instantons. Given the close relationship of the
biKaehler topological sigma model and the $A$ model, it is natural to expect 
the biKaehler model to be also characterized by a rich structure of topological 
sectors and world sheet instantons. A generalization of quantum cohomology 
would emerge in this way. 

Since biKaehler geometry allows one to construct a large number of topological
terms to be added to the sigma model action by hand, the range of possibilities in
which the quantum deformation could be carried out in the biKaehler model is 
far wider than that of the $A$ model. So, as a preliminary step, it seems that
a classification of the meaningful topological terms would be required. 

{\it A biKaehler sigma model containing the $B$ model}

The authors of ref. \cite{AKSZ} were able to obtain both the $A$ and $B$
topological sigma models by gauge fixing suitable actions 
satisfying the Batalin--Vilkovisky master equation associated with the
appropriate odd symplectic form. 
While the $A$ model has appeared in our analysis as a particular case of the
biKaehler model, the $B$ model has been conspicuously absent.
Presumably, the $B$ model can be obtained by gauge fixing the generalized Kaehler
Hitchin model action $S_{GK}$ in a way analogous to that followed in sect. \ref{sec:Hitchin}.
We have not been able to do that so far, due to our present limited 
understanding reality conditions of the fields and 
of the geometry of the appropriate field space Lagrangian 
submanifold $\mathfrak{L}$. This is definitely an issue calling for 
further investigation. 

\vskip.5cm
\begin{small}
\section*{Acknowledgments}
\end{small}

We are grateful to T. Strobl for introducing and explaining to us 
the importance of the work of ref. \cite{AKSZ}. We are also grateful to
F. Bastianelli and G. Velo for steady encouragement. 

\vfill\eject

\end{document}